\documentclass[12pt]{article}

\usepackage{epsfig}

\usepackage{amssymb}
\usepackage{amsfonts}

\usepackage{color}
\def\be{\begin{equation}}
\def\ee{\end{equation}}
\def\ba{\begin{array}{c}}
\def\ea{\end{array}}

\def\ben{$$}
\def\een{$$}

\newcommand{\bea}{\begin{eqnarray}}
\newcommand{\eea}{\end{eqnarray}}

\newcommand{\kt}{\rangle}
\newcommand{\br}{\langle}

\newtheorem{thm}{Theorem}

\newtheorem{lemma}[thm]{Lemma}

\newtheorem{conj}[thm]{Conjecture}

\begin{document}
%
%\textcolor{black}{xxx}

%. \vspace{1cm}

 \begin{center}{\Large \bf

Inverse Feshbach's problem: Solvability and solutions

  }

\vspace{0.8cm}

  {\bf Miloslav Znojil}$^{a,b,c,d,}$\footnote{{e-mail:
znojil@ujf.cas.cz}}

\end{center}

\vspace{10mm}

 $^{a}${The Czech Academy of Sciences,
 Nuclear Physics Institute, \v{R}e\v{z} 292,
250 68 Husinec,\\. \hspace{1cm} Czech Republic}

%{e-mail: znojil@ujf.cas.cz}}

 $^{b}${Department of Physics, Faculty of
Science, University of Hradec Kr\'{a}lov\'{e}, Rokitansk\'{e}ho 62,
\\. \hspace{1cm} 500 03 Hradec Kr\'{a}lov\'{e},
 Czech Republic}

$^{c}${School for Data Science and Computational Thinking,
Stellenbosch University, \\. \hspace{1cm} 7600 Stellenbosch,
 South Africa}
%Research Fellow

 $^{d}${Institute of System Science, Durban University of Technology,
4001 Durban, South Africa}

\vspace{5mm}

%\newpage

\section*{Abstract}

The well known Feshbach-inspired isospectral
replacement of a
conventional quantum Hamiltonian $H$
by its energy-dependent but user-friendlier
effective-Hamiltonian alternative
$H^{}_{e\!f\!f}(E)$ (acting in a suitable subspace) is a
more or less routine procedure.
In contrast,
the
inverse problem of
reconstruction $H^{}_{e\!f\!f}(E) \to H$
is considered unfeasible.
Recently,
the skepticism has been further enhanced in
Phys. Lett. A 556 (2025) 130816 where
we argued that
the
reconstruction
remains prohibitively difficult
even after a maximal simplification of
the target matrix $H$
(assumed to be partially tridiagonalized).
In the present paper
the latter skepticism is disproved.
It is conjectured
(and, for a number of special cases,
it is constructively demonstrated)
that
the ``missing''  matrix elements of $H$
can be reconstructed and defined,
in terms of the relevant matrix elements
of $H^{}_{e\!f\!f}(E)$, in
closed form.

%\newpage

\section*{Keywords}.

model space;

effective Hamiltonians;

inverse problem of reconstruction of
a full-space quantum Hamiltonian;

coupled set of polynomial algebraic equations;

explicit and recursive exact solutions;

%
%\newpage
%
%\subsection*{Highlights}
%
%.
%
%measured energy levels assumed fitted by an effective Hamiltonian $H^{}_{e\!f\!f}(E)$;
%
%the full-space Hamiltonian $H$ assumed tridiagonal out of the model space;
%
%inverse-problem reconstruction  $H^{}_{e\!f\!f}(E) \to H$ reduced to a
%set of nonlinear equations;
%
%multiple closed-form solutions constructed;
%

\newpage

\section{Introduction\label{introduction}}

In a broad range of applications of quantum mechanics our
choice of the so-called model-space projector
 \be
 P=\sum_{m=1}^{M}|m\kt \br m|
 \,
 \label{enproj}
 \ee
plays a key role because it
enables us to replace a preselected Hamiltonian operator $H$ by its
truncated, variationally motivated
$M$-by-$M$-matrix approximation,
 \be
 H\ \to \
 PHP=H^{(M)}=
  \left[ \begin {array}{cccc}
  H_{11}&H_{12}&\ldots&H_{1M}
   \\
  H_{21}&H_{22}&\ldots&H_{2M}
   \\
 \vdots
 &\vdots&\ddots&\vdots
 \\
  H_{M1}&H_{M2}&\ldots&H_{MM}
   \\
   \ea
   \right ]\,.
 \label{trunca}
   \ee
Such a finite-dimensional
matrix can be diagonalized by the standard
linear-algebraic techniques yielding the
approximate
spectrum of bound-state energies which is
expected to fit the measured
experimental data.

Whenever we employ a sufficiently large cut-off
$M$, the variational mathematical background of the evaluated
energies would make them almost precise. In such a case
our initial choice of the
Hamiltonian can be declared realistic
if and only if the quality
of the fit of the data appears sufficiently persuasive.

In practice, unfortunately, we are often forced to work with $M$
which is not sufficiently large. Then, we have to test also the reliability
of the approximation. Such a test can be performed, e.g., via a
repetition of the diagonalization after an enlargement of the
dimension, $M \to M+K$. A disadvantage is that the costs of the
iterated diagonalization might grow quickly with $K$, especially
when these costs are dominated by the evaluation of the new matrix
elements themselves.

The variational technique is to be complemented, in such a
situation, by perturbation theory.
Typically, people follow the Feshbach's idea
\cite{Feshbach} and replace the truncated matrix $H^{(M)}$ by its
``effective'' $M$-by-$M$-matrix amendment $H^{(M)}_{e\!f\!f}$. The
influence of the out-of-the-model-space matrix elements of
$H$ becomes simulated by a manifest
energy-dependence of at least some of the matrix elements of
$H^{(M)}_{e\!f\!f}=H^{(M)}_{e\!f\!f}(E)$ \cite{Lomb}.

The replacement $H^{(M)} \to H^{(M)}_{e\!f\!f}(E)$
(see its more detailed
description in section \ref{2introduction} below)
found
a number of successful applications ranging from the Feshbach-inspired
approaches to nuclear theory \cite{Feshbachb,Feshbachc}
up to the L\"{o}wdin-inspired
use of the isospectrality between $H$ and $H^{(M)}_{e\!f\!f}(E)$
in quantum chemistry \cite{Loewdin,Baha},
in the so called PT-symmetric quantum models \cite{PT},
in non-commutative quantum mechanics \cite{neco},
in relativistic quantum physics \cite{DKP}
or even in information theory \cite{info}.
In mathematics, of course, the replacement
can be given a mathematically entirely rigorous ground
(see, e.g., Refs.~\cite{Hill,Hillb}).
Still,
even an
{\em approximate\,} reconstruction of
$H^{(M)}_{e\!f\!f}(E)$ from its full-space partner $H$
has often been found sufficient
for practical purposes
(see, e.g., Refs.~\cite{Mares,Loewdinb}).

The efficiency of the latter, approximative model-building strategy
proved so persuasive that people started using the energy-dependent
candidates for the effective Hamiltonian even without any reference
to the full-space $H$ (cf.,
e.g., \cite{Maresb,Maresc,Marescc,Maresd,Marese}).
In our recent critical letter \cite{PLAI} we
pointed out, therefore, that a
successful fit of the available experimental data using an
energy-dependent {\it ad hoc\,} matrix $H^{(M)}_{e\!f\!f}(E)$
{\em need not\,} necessarily make the model compatible with the basic
postulates of quantum mechanics.

We imagined that multiple {\it ad hoc\,} matrices
$H^{(M)}_{e\!f\!f}(E)$ simply need not possess {\em any\,}
fundamental full-space operator partner $H$ at all.
In general, indeed,
the necessary proof of the existence of $H$ would be
difficult.
In \cite{PLAI} we proposed, therefore, that
it makes sense to accept
certain auxiliary technical
constraints, under which
the necessary
confirmation of internal consistency of the
models using effective-Hamiltonian
would
still become feasible.

For an explicit illustration of the recipe
we choose a rather elementary model in which
the reconstruction of the unknown full-space
Hamiltonian $H^{(N)}$ from its known
energy-dependent partner
$H^{(M)}_{e\!f\!f}(E)$
has been performed
at
the first nontrivial choice of $N=M+1$.
Naturally, such a sample of the implementation of our
innovative methodical
idea was not too persuasive.
For this reason we decided to turn attention, more recently,
to the study of the less academic reconstruction
scenarios in which
the differences
$K=N-M $ between the
respective dimensions of the
``full'' Hilbert space (assumed, still, finite-dimensional)
and its model-space subspace
would not be restricted to the first few smallest integers.

In this respect, incidentally,
we still
remained skeptical in \cite{PLAI}. We felt
that in spite of the underlying methodical innovation,
the practical
implementations
of the reconstruction seemed to remain
difficult. Requiring, among others,
an ample use
of certain dedicated, computer-mediated
algebraic symbolic manipulations, etc.
In this context it is possible
to re-emphasize that our present paper can be read
as an immediate constructive
extension and continuation of the project.
Concerning the
task of the explicit inverse-problem reconstruction
  \be
 H^{(M)}_{e\!f\!f}(E)\ \to \ H^{(M+K)}\,,
 \ \ \ \  K < \infty\,,
 \label{praka}
  \ee
our paper will announce
a decisive conceptual as well as technical progress.

The overall strategy of the reconstruction
will be recalled and outlined in
section~\ref{3introduction}. We will emphasize there that
the abstract conceptual innovation
as presented in letter \cite{PLAI}
has to be complemented
by a more explicit analysis of its
constructive aspects.

We will
describe, first of all, the sources of
our previous skepticism.
Their essence will be
reanalyzed
in section \ref{4introduction}.
We will show there that the decisive technical
difficulties already emerge as early as
at $K=3$. Even though
the reconstruction of $ H^{(M+3)}$
remains
feasible
in technical terms
(i.e., it remains
fully non-numerical and
keeps having an explicit algebraic
form),
its main weakness has been found to lie in an
excessive
length of the ultimate
formulae.

We will show that
after a subtle change of the approach
the practical applicability of the inverse-problem
reconstruction (\ref{praka})
can still be newly re-evaluated as
satisfactory.
This claim will be supported by two new results.
Firstly, in sections \ref{5introduction}, \ref{5Bintroduction}
and  \ref{5Cintroduction} we will show that
the definition of the reconstructed $H^{(M+K)}$
can be made, at \textcolor{black}{quite a few} finite values of $K$,
compact. 

In the light of what has just been said,
this may sound like a paradox,
but its resolution is elementary,
based on a transition from the entirely explicit
definitions of the individual
reconstructed matrix elements $H^{(M+K)}_{jk}$
as functions of the set of the ``input'' matrix elements
of $H^{(M)}_{e\!f\!f}(E)$
to their more sophisticated
recursive-evaluation version.
A key merit of such a result may be seen in the
discovery of the existence of the recursion-specifying formulae
which acquire a ``stabilized''
canonical form at any $K \geq K_{stab} $
where the latter bound of complexity $K_{stab}$ remains
element-subscript-dependent,
$K_{stab}=K_{stab}(j,k)$.

Incidentally,
the bound $K_{stab}$
appeared not to be always sufficiently sharp.
This opened
a few new
challenging questions,
a partial answer to which will be
outlined
in our last three sections~\ref{6introduction}, \ref{discussion} and \ref{susu}.

\section{What is precisely the effective Hamiltonian?\label{2introduction}}

\subsection{The concept of model space\label{duction}}

The introduction of projector
(\ref{enproj}) and of its
complement $Q=I-P$ enables us to partition the full-space Hamiltonian
 \be
 H_{}= \left[ \begin {array}{cc}
  PH_{}P&PH_{}Q\\
  QH_{}P&QH_{}  Q
 \end {array} \right]
 \,
 \label{pafukit}
 \ee
as well as
Schr\"{o}dinger equation,
 \be
 \left (
 \begin{array}{cc}
 P(H_{}-E^{(n)})P&PH_{}Q\\
 QH_{}P&Q(H_{}-E^{(n)})Q
 \end{array}
 \right )\,
 \left (
 \begin{array}{cc}
 P\,|\psi^{(n)}\kt\\
 Q\,|\psi^{(n)}\kt
 \end{array}
 \right )=0\,,
 \ \ \ \ n = 0,1,  \ldots
 \,.
 \label{[u2]}
 \ee
This opens the way towards the
separation
of the out-of-the-model-space component
of the wave function,
 \be
 Q\,|\psi^{(n)}\kt=
 -Q\,\frac{1}{QH_{} Q-E^{(n)}}\,Q\,H_{}\,|\phi^{(n)}\kt
 %\,,
 %\ \ \ \ \
 \,
 \label{[3]}
 \ee
where we abbreviated
 $|\phi^{(n)}\kt=
 P\,|\psi^{(n)}\kt$.
The insertion of expression (\ref{[3]})
then reduces our full-space Schr\"{o}dinger Eq.~(\ref{[u2]})
to its strictly isospectral model-space-projected
version
 \be
 H_{e\!f\!f}(\eta)\,|\phi^{(n)}\kt = E^{(n)}\,|\phi^{(n)}\kt
 \,,\ \ \ \ \eta=E^{(n)}
 \,,
 \ \ \ \ \ n = 0, 1, \,\ldots \,
 \label{ethee}
 \ee
containing the energy-dependent $M$-by-$M$-matrix effective
Hamiltonian
 \be
 H_{\!\it ef\!f}(E)=H_{\!\it ef\!f}^{(M)}(E)=
 P \,H_{}\,P
 - P\,H_{}\,Q\,\frac{Q}{Q\,H_{}\,Q-E}\,Q\,H_{}\,P
 \,.
 \label{formu}
 \ee
The
construction is motivated by the observation that the original
ket-vector (of the system living in an $N-$dimensional Hilbert space
$ {\cal H}$, finite or infinite, i.e., with $N=M+K$ and $K \leq \infty$)
can be split into its
model-space
part and the $Q-$projected component of Eq.~(\ref{[3]})
with a weaker influence, say, on the most relevant
low-lying part of the spectrum.

%\subsection{Effective Schr\"{o}dinger equation}

\subsection{Reduction of $H$
to $H_{e\!f\!f}(\eta)$\label{edea}}

The smallness of
influence of the matrix elements $H_{mn}$ with $m>M$ and/or $n>M$
is an important assumption which
makes the choice of the
basis in the less relevant, $Q-$projected part of the full
Hilbert space
more flexible.
In \cite{PLAI} we
recommended,
therefore, the use of such a basis which would
render the
out-of-the-model-space \textcolor{black}{$K$-by-$K$-matrix}
part 
\textcolor{black}{$QHQ$}
of the Hamiltonian matrix $H=H^{(N)}$
tridiagonal. \textcolor{black}{Moreover,
using }
another, slightly larger projector
 \be
 Q^{(+)}=|M\kt \br M|+Q
 =\sum_{m=M}^{M+K}|m\kt \br m|\,,
 \ \ \ \ \ K \leq \infty
 %\,
 \label{qenproj}
 \ee
\textcolor{black}{we decided to extend our more or less
purely technical assumption of the
``user-friendly''
tridiagonality to the slightly larger matrix 
$Q^{(+)}HQ^{(+)}$.}

\textcolor{black}{In this manner we obtained the following, ultimate
form of our
assumption specifying the class of the full-space Hamiltonians
of our present interest},
 \be
 H=
  \left[ \begin {array}{ccc|c|ccccc}
  H_{1\,1}&\ldots
 &H_{1\,M-1}&H_{1\,M}&0&0&\ldots&\ldots&0
   \\
 \vdots&
 &\vdots&\vdots&\vdots&\vdots &&
 &\vdots
    \\
 H_{M-1\,1}&\ldots
 &H_{M-1M-1}&H_{M-1M}&0&0&\ldots&\ldots
 &0
    \\
 \hline
 H_{M\,1}&\ldots
 &H_{MM-1}&H_{MM}&H_{MM+1}&0&\ldots
 &\ldots&0
   \\
 \hline
     0&\ldots
 &0&H_{M+1M}&H_{M+1M+1}&H_{M+1M+2}&0
 &\ldots&0
   \\
   0&\ldots
 &0
 &0&H_{M+2M+1}&H_{M+2M+2}&\ddots&\ddots
 &\vdots
   \\
 \vdots&&\vdots &\vdots&0
 &\ddots&\ddots&H_{N-2N-1}&0
   \\
 \vdots&&\vdots &\vdots&\vdots&\ddots&H_{N-1N-2}&H_{N-1N-1}&H_{N-1N}
    \\
  0&\ldots&0&0&0&\ldots&0&H_{NN-1}&H_{NN}
    \\
 \end {array} \right].
 \label{hejkitie}
 \ee
\textcolor{black}{In it, the triply-partitioned basis-specifying}
constraint
\textcolor{black}{can be re-read as a very natural
``doorway-state-mediated decoupling''
of the subspaces (see also a more detailed 
discussion of the physics behind such a ``doorway-state-interaction'' 
hypothesis
in \cite{PLAI}).}

\textcolor{black}{
Simultaneously, the latter hypothesis
{\it  alias\,} ``enhanced tridiagonality'' assumption has also}
been found to lead to several \textcolor{black}{important}
technical
simplifications.
\textcolor{black}{Mathematically,
they were independently supported and inspired, after all, 
by the existing and truly extensive
literature on the tridiagonal
(or, at least, block-tridiagonal)
finite or infinite 
and/or Hermitian or non-Hermitian matrices
(cf., e.g., papers \cite{CF,Has,Ha,Hashim,Hash,PS1}
as just a very small sample of the related relevant references
and results).}

\textcolor{black}{{\it Pars pro toto\,} let us mention here,
{\it expressis verbis,} that at} 
any $N=M+K$, indeed,
the
requirement of the isospectrality between Schr\"{o}dinger
Eqs.~(\ref{[u2]}) and
(\ref{ethee}) immediately implies that
the related
effective Schr\"{o}dinger
operator becomes particularly elementary,
having the form
 \be
 H_{\!\it ef\!f}-E=
  \left[ \begin {array}{cccc}
  H_{11}-E&\ldots
 &H_{1M-1}&H_{1M}
   \\
 \vdots&\ddots
 &\vdots&\vdots
    \\
 H_{M-11}&\ldots
 &H_{M-1M-1}-E&H_{M-1M}
    \\
 H_{M1}&\ldots
 &H_{MM-1}&{\cal G}(E)
   \\
 \end {array} \right]\,
 \label{finkit}
 \ee
in which the difference from its truncated-matrix
predecessor remains restricted,
in the light of Eq.~(\ref{formu}),
to the single matrix element
 \be
 {\cal G}(E)=
 H_{MM}-E
 -H_{MM+1}\,
 \left [\frac{Q}{Q\,H\,Q-E}
 \right ]_{M+1M+1}\!\!\!\!\!\!H_{M+1M}\,.
 \label{rfund}
 \ee
The use of 
\textcolor{black}{Eq.~(\ref{qenproj})}
enables us finally to arrive at the definition
 \be
 {\cal G}(E)=1/f_0(E)\,,\ \ \ \
 f_0=
 \left [\frac{I}{Q^{(+)}\,H\,Q^{(+)}-E}
 \right ]_{MM}\,
 \label{urfund}
 \ee
which is amazingly compact (cf. also
equation Nr. 28 in \cite{PLAI}
\textcolor{black}{which offers
an explicit mathematical continued-fraction 
representation and 
interpretation of this 
function of energy -- in this respect,
it is also worth mentioning that 
the authors of paper \cite{Has}
would call this expression the
``m-function of a
semi-infinite Jacobi matrix $Q^{(+)}\,H\,Q^{(+)}$''}).

\subsection{Example}

For an illustration
of the construction of $H_{\!\it ef\!f}(E)$
let us choose
the first nontrivial value of $K=1$, i.e.,
let us consider
the full-space
$(M+1)$-by-$(M+1)$-matrix input Hamiltonian
of the following form,
 \be
 H^{(M+1)}=
  \left[ \begin {array}{ccc|c|c}
  H_{11}&\ldots
 &H_{1M-1}&H_{1M}&0
   \\
 \vdots&
 &\vdots&\vdots
 & \vdots
    \\
 H_{M-11}&\ldots
 &H_{M-1M-1}&H_{M-1M}&0
    \\
 \hline
 H_{M1}&\ldots
 &H_{MM-1}&H_{MM}&H_{MM+1}
   \\
 \hline
     0&\ldots
 &0&H_{M+1M}&H_{M+1M+1}
   \\
 \end {array} \right]\,.
 \label{1foukitie}
 \ee
The hypothetical full Hilbert space
would be then just $(M+1)-$dimensional,
$ {\cal H}={\cal H}^{(M+1)}$, being
spanned by such a suitably adapted
basis that the $(2M-2)-$plet of the outermost matrix
elements of the Hamiltonian are strictly equal to zero:
Such a goal can be
achieved via
a sequence of
the so-called Jacobi rotations.

At any $K$, our effective Hamiltonian
(\ref{finkit})
only
differs from
its truncated-matrix predecessor $H^{(M)}$
in
a single matrix element,
with $H^{(M)}_{MM}$
replaced by $\left [H_{\!\it ef\!f}(E)
\right ]_{MM}= {\cal G}(E)+E$.
At $K=1$,
in particular,
such an energy-dependent function is particularly simple,
 \be
 {\cal G}(E)=
 H_{MM}-E
 -H_{MM+1}\,H_{M+1M}\,
 /\left ({H_{M+1M+1}-E}
 \right )\,,
 \ \ \ \ K=1\,.
 \label{vbrfund}
 \ee
Nevertheless, even a cursory
inspection of this formula
reveals how thoroughly the partial tridiagonalization of $H$ simplifies
the effective Hamiltonian (\ref{finkit}).

The single energy-dependent matrix element ${\cal G}(E)$
of the
effective Hamiltonian
need not necessarily
be a
smooth function of $E$.
Thus,
as a consequence,
even the direct-problem process of solution of
the effective Schr\"{o}dinger Eq.~(\ref{ethee})
will depend strongly on the properties of this function.
During the process of solution
of the effective Schr\"{o}dinger equation
we have to proceed, in general, iteratively,
Thus, we have to pick up a tentative value of $\eta$
and try to reach the convergence $\eta \to E^{(n)}$,
the rate of which
need not necessarily be always quick.

Such an observation also stands behind
our present project
in which we will try to
enlarge the model-space dimension
via a tentative inversion (\ref{praka})
of the reduction  of $H$
to $H_{e\!f\!f}(\eta)$.

%: Deduction of $H$
%from $H_{e\!f\!f}(\eta)$

\section{Inverse problem\label{3introduction}}

In \cite{PLAI} we managed to
reduce the
reconstruction
of
the ``fundamental'', full-space $N$ by $N$ matrix
$H=H^{(N)}$
with $N = M+K$ to the solution of a
nonlinear algebraic
set of $2K+1$ coupled polynomial equations.
For illustration we choose
the model with $K=1$, and
we admitted that
at any larger $K$ the construction of
the solution $H^{(N)}$
would be a nontrivial
task.

In our present paper,
we intend to show,
that -- and how -- the general $K>1$ Hamiltonian-reconstruction
problem can systematically be solved in closed form.
\textcolor{black}{Naturally, multiple versions of
such an inverse problem
can be found presented and discussed in the related literature
(cf., first of all,
a truly exhaustive coverage of
this problem in the review-like papers \cite{Has} and \cite{Ha}).
In this context, 
a characteristic 
specific feature of our present approach
(as proposed in \cite{PLAI})
is that in contrast to the majority of conventional approaches
we will not need to know,
for the purposes of our reconstruction, the 
large-energy asymptotics
of 
our ``dynamical information carrying''
m-function ${\cal G}(E)$.}

%\newpage the partially tridiagonalized Hamiltonian

\subsection{\textcolor{black}{The question of uniqueness of the}
reconstruction of $H^{(M+K)}$
\label{idea}}

Our goal is an explicit, constructive  assignment of a large
and partially tridiagonalized
matrix $H=H^{(N)}$
of Eq.~(\ref{hejkitie})
to a preselected effective Hamiltonian $H_{e\!f\!f}^{(M)}(E)$.
We intend to use the method of our letter
\cite{PLAI} in which we postulated that
we know, in advance, only the effective Schr\"{o}dinger
operator $H^{(M)}_{e\!f\!f}(E)-E$ of Eq.~(\ref{finkit})
or, in other words, that
we only know its single energy-dependent
matrix element ${\cal G}(E)$.

Our task is to reconstruct all of the not yet known
matrix elements of $H^{(M+K)}$.
For the sake of clarity we may amend the notation
and denote
 \be
 H^{(N)}=%
% H^{(M+K)}=
  \left[ \begin {array}{ccc|c|ccccc}
  H_{11}&\ldots
 &H_{1M-1}&H_{1M}&0&0&\ldots&\ldots&0
   \\
 \vdots&
 &\vdots&\vdots&\vdots&\vdots &&
 &\vdots
    \\
 H_{M-11}&\ldots
 &H_{M-1M-1}&H_{M-1M}&0&0&\ldots&\ldots
 &0
    \\
 \hline
 H_{M1}&\ldots
 &H_{MM-1}&a_0&b_0&0&\ldots
 &\ldots&0
   \\
 \hline
     0&\ldots
 &0&c_1&a_1&b_1&0
 &\ldots&0
   \\
   0&\ldots
 &0
 &0&c_2&a_2&b_2&\ddots
 &\vdots
   \\
 \vdots&&\vdots &\vdots&0
 &\ddots&\ddots&\ddots&0
   \\
 \vdots&&\vdots &\vdots&\vdots&\ddots&c_{K-1}&a_{K-1}&b_{K-1}
    \\
  0&\ldots&0&0&0&\ldots&0&c_{K}&a_{K}
    \\
 \end {array} \right]\,
 \label{kitie}
 \ee
where all of the to-be-reconstructed
matrix elements of $H^{(N)}$ are denoted
by the lower-case symbols
(cf. also equation Nr. 23 in \cite{PLAI}).

\textcolor{black}{In order to avoid misunderstandings,
it is immediately necessary to emphasize here that }
for the reasons
which may be found explained in \cite{PLAI},
\textcolor{black}{matrix (\ref{kitie}) {\em cannnot\,}
in fact be
the result of inversion $H^{(M)}_{e\!f\!f}(E)\ \to \ H^{(M+K)}$
of Eq.~(\ref{praka})
because in the light of the explicit continued-fraction
definition of ${\cal G}(E)$ (given, e.g., by equation nr. 12 
in \cite{PLAI}), the lower-case matrix elements $b_n$ and $c_{n+1}$
of matrix (\ref{kitie})
only enter this 
continued-fraction
definition 
in the form of their product 
$b_n\,c_{n+1}$ to be denoted by a dedicated symbol $\rho_n$.}

\textcolor{black}{For this reason we have to
decide to treat, say, the 
numbers $c_1, c_2,\ldots,c_k$
filling the lower diagonal in (\ref{kitie})
as a $K-$plet of free parameters.
Keeping this ``last reformulation'' of our inverse problem
in mind, we can 
reformulate it (and make its solution, finally, unique)
when we finally} 
modify 
\textcolor{black}{our ansatz for}
the 
latter Hamiltonian,
and when we replace it
by its exactly isospectral partner
 \be
 H=
 H^{(N)}=
 %\left[ \begin {array}{cc}
%  SH_{}S&SH_{}L\\
%  LH_{}S&LH_{}  L
% \end {array} \right]=
  \left[ \begin {array}{ccc|c|ccccc}
  H_{11}&\ldots
 &H_{1M-1}&H_{1M}&0&\ldots&&\ldots&0
   \\
 \vdots&
 &\vdots&\vdots&\vdots&&&
 &\vdots
    \\
 H_{M-11}&\ldots
 &H_{M-1M-1}&H_{M-1M}&0&\ldots&&\ldots
 &0
    \\
 \hline
 H_{M1}&\ldots
 &H_{MM-1}&a_0&\rho_0&0&\ldots
 &\ldots&0
   \\
 \hline
     0&\ldots
 &0&1&a_1&\rho_1&0
 &\ldots&0
   \\
   0&\ldots
 &\ldots
 &0&1&a_2&\ddots&\ddots
 &\vdots
   \\
 \vdots&&&\vdots&0
 &1&\ddots&\rho_{K-2}&0
   \\
 \vdots&&&\vdots&\vdots&\ddots&\ddots&a_{K-1}&\rho_{K-1}
    \\
  0&\ldots&\ldots&0&0&\ldots&0&1&a_{K}
    \\
 \end {array} \right]\,
 \label{fkitie}
 \ee
We abbreviated $b_m c_{m+1}:=\rho_m$, having still
guaranteed the isospectrality --
cf. also the same \textcolor{black}{fundamental
initial reconstruction ansatz as given by} equation Nr. 24 in \cite{PLAI}.

The main idea supporting the possibility of reconstruction
of all of the lower-case elements of the
latter matrix (i.e., of its \textcolor{black}{to be reconstructed} submatrix \textcolor{black}{
 $$
 {\cal S}^{(K+1)}=Q^{(+)}\,H\,Q^{(+)}
 =
   \left[ \begin {array}{cccccc}
 a_0&\rho_0&0&\ldots
 &\ldots&0
   \\
    1&a_1&\rho_1&0
 &\ldots&0
   \\
   0&1&a_2&\ddots&\ddots
 &\vdots
   \\
 \vdots&0
 &1&\ddots&\rho_{K-2}&0
   \\
 \vdots&\vdots&\ddots&\ddots&a_{K-1}&\rho_{K-1}
    \\
  0&0&\ldots&0&1&a_{K}
    \\
 \end {array} \right]\,
 $$}
as defined also by equation Nr.~25 in \cite{PLAI})
lies in our basic assumption that we know all of
the values of the real function ${\cal G}(E)$
in a certain empirically specified
interval of energies $E \in (E_{\min},E_{\max})$.

This means that at an arbitrary
element $E_\alpha$ of this interval
we can treat Eq.~(\ref{urfund})
as a mathematical identity.
Thus, its left-hand-side value ${\cal G}(E_{\alpha})$
is, by our assumption, known at any non-degenerate
set of \textcolor{black}{energies with} subscripts, say,
$\alpha=0,1, \ldots\,,{Z}$.
As long as their number can be chosen equal to the
number of the unknown matrix elements (i.e., once we put $Z = 2K$),
and as long as the left-hand-side component of the identity~(\ref{urfund})
is just the ratio of two polynomial functions of these elements,
we arrive at a new interpretation of the identity
which can serve as a set of  $2K+1$
nonlinear algebraic equations
\be
 {\cal G}(E_{\alpha})=1/f_0(E_{\alpha})\,,\ \ \ \
 \alpha=0,1, \ldots\,,2K\,.
 \label{eurfund}
 \ee
They define,
in implicit manner, the whole
initially unknown matrix
${\cal S}^{(K+1)}$
(cf. also the last general equation Nr.~29 in \cite{PLAI}).

The rest of our present paper will be now devoted to
the ways of making such an implicit definition explicit,
\textcolor{black}{without recalling the standard mathematical
spectral-analytic
theory \cite{Has,PS1} in which one has to work with the 
$E \to \infty$ asymptotics of 
the Weyl's m-function ${\cal G}(E)$}.

\subsection{The method and explicit construction at $K=1$ \label{spa9}}

At $K=1$,
the
``unknown'', to-be-reconstructed
matrix elements of $H=H^{(M+1)}$ are just four, viz.,
 $$H_{MM}:=a_0\,,H_{MM+1}=b_0\,,H_{M+1M}=c_1\,,H_{M+1M+1}=a_1\,.$$
As long as they are to be deduced from our ``input'' knowledge of
the effective Hamiltonian,
we denoted them by the dedicated lower-case symbols.
With 
\textcolor{black}{$H_{M+1M}=c_1$ kept as an arbitrary
optional parameter, and with}
the to-be-reconstructed
triply partitioned full-space Hamiltonian
 \be
 H^{(M+1)}=
  \left[ \begin {array}{ccc|c|c}
  H_{11}&\ldots
 &H_{1M-1}&H_{1M}&0
   \\
 \vdots&
 &\vdots&\vdots
 & \vdots
    \\
 H_{M-11}&\ldots
 &H_{M-1M-1}&H_{M-1M}&0
    \\
 \hline
 H_{M1}&\ldots
 &H_{MM-1}&a_0&\rho_0
   \\
 \hline
     0&\ldots
 &0&1&a_1
   \\
 \end {array} \right]\,
 \label{1bfoukitie}
 \ee
just the evaluation of the three lower-case matrix
elements is needed.
Thus,
we have
to determine
the matrix
 \be
 {\cal S}_{}^{(2)}=
  \left[ \begin {array}{cc}
  a_0&\rho_0
   \\
     1&a_1
      \\
 \end {array} \right]\,
 \label{1spodkit}
 \ee
using equations
(\ref{eurfund}), i.e., using the triplet of relations
 \be
 {E_{\alpha}}^{2}+G_{\alpha} \, E_{\alpha}
  -  E_{\alpha}\,\left( a_{{0}}+a_{{1}} \right)
  -G_{\alpha}\,a_{{1}}
  +a_{{0}}
 a_{{1}}-{\it \rho}_{{0}}=0\,,\ \ \ \ \alpha=0,1,2\,.
 \label{ome}
 \ee
These equations can be rewritten
as a system
 \be
 G_{\alpha}\,y_{{1}}
 -E_{\alpha}\,x_{{1}}
 -x_{{2}}=
 {E_{\alpha}}^{2}+G_{\alpha}  E_{\alpha}\,,\ \ \ \ \alpha=0,1,2\,.
 \label{epitoma}
 \ee
which is linear in the three
new unknown parameters defined as follows,
 \be
  x_{{1}}=-a_{{0}}-a_{{1}}\,,
  \ \ \
  x_{{2}}=a_{{0}}a_{{1}}-{\it \rho}_{{0}}\,,\  \ \
  y_{{1}}=a_{{1}}
 \,.
 \label{toma}
 \ee
The system (\ref{epitoma})
is solvable by
a very routine matrix inversion so that the
whole process of reconstruction of $H$
becomes reduced to the
only nontrivial task of
inversion of the
mapping of Eq.~(\ref{toma}).
At $K=1$, even this task is also easy:

\begin{lemma}
\label{lemmakje1}
At $K=1$ the inversion of
 the mapping
 $\{a_0,a_1,\rho_0\} \to \{x_1,x_2,y_1\}$
is
given by formulae
 \be
  a_{{1}}=y_{{1}}\,,\ \ \ \
 a_{{0}}=-x_{{1}}-y_{{1}}\,,\ \ \ \
 {\it \rho}_{{0}}=-x_{{1}}y_{{1}}-x_{{2}}-{y_{{1}}}^{2} \,.
 \label{epi}
 \ee
\end{lemma}

 \noindent
This is the result which we
used as an illustrative example in our
methodical letter \cite{PLAI}.
At the same time, we did not accompany it by
any of its $K>1$
descendants -- for
the reasons which will be explained in the next section.

%(\ref{finkit}) solvability

%%\newpage

%

\section{Solvability
beyond $K=1$\label{4introduction}}

In our
notation
the set of the matrix elements of
the target Hamiltonian $H^{(M+K)}$ in Eq.~(\ref{praka})
is split
into its known (i.e., upper-case) matrix elements
and its unknown and
to-be-reconstructed (i.e., lower-case)
components (cf. Eq.~(\ref{fkitie}) above).
As we already mentioned,
the project of the reconstruction of the latter set
is well motivated
by the needs of the experimental quantum physics
in which
a
measured multiplet
$E^{(1)}$, $E^{(2)}$, \ldots, $E^{(N)}$
of the empirical bound-state energies
is often
well
fitted
using just
an {\it ad hoc\,} model-space Schr\"{o}dinger Eq.~(\ref{ethee}).

\subsection{Reconstruction in implicit form}

%{Gr\"{o}bner basis solutions\label{spa92} }

Under our partial-tridiagonality assumptions concerning $H^{(M+K)}$,
all of the necessary
input information about dynamics
is carried, exclusively, by the single
matrix element
$$\left [H_{\!\it ef\!f}\right ]_{MM}(E)= {\cal G}(E)+E$$
of our effective Hamiltonian.
A formal key to the reconstruction of  $H^{(M+K)}$
lies then in the free variability of $E$.
The point is that
function ${\cal G}(E)$ can be also perceived, alternatively
(i.e., via Eqs.~(\ref{rfund}) or~(\ref{urfund})), as a well-defined
rational function of all of
the unknown and missing lower-case matrix elements of $H^{(M+K)}$.

In a preparatory step let us restrict our attention to the
two alternative
auxiliary forms of the to-be-specified submatrices of $H^{(M+K)}$, viz.,
 \be
 Q^{(+)}\,H\,Q^{(+)}=
  \left[ \begin {array}{cccccc}
  a_0&\rho_0&0&\ldots
 &\ldots&0
   \\
     1&a_1&\rho_1&0
 &\ldots&0
   \\
   0&1&a_2&\rho_2&\ddots
 &\vdots
   \\
 0&\ddots
 &\ddots&\ddots&\ddots&0
   \\
 \vdots&\ddots&0&1&a_{K-1}&\rho_{K-1}
    \\
  0&\ldots&0&0&1&a_{K}
    \\
 \end {array} \right]:={\cal S}^{(K+1)}\,
 \label{spodkit}
 \ee
and
 \be
 Q^{}\,H\,Q^{}=
  \left[ \begin {array}{cccccc}
  a_1&\rho_1&0&\ldots
 &\ldots&0
   \\
     1&a_2&\rho_2&0
 &\ldots&0
   \\
   0&1&a_3&\rho_3&\ddots
 &\vdots
   \\
 0&\ddots
 &\ddots&\ddots&\ddots&0
   \\
 \vdots&\ddots&0&1&a_{K-1}&\rho_{K-1}
    \\
  0&\ldots&0&0&1&a_{K}
    \\
 \end {array} \right]:={\cal R}^{(K)}
 \,.
 \label{rpodkit}
 \ee
This enables us to deduce that
 \be
 f_0=f_0(E)=\frac{\det \left ({\cal R}^{(K)}-E\right )}
 {\det \left ( {\cal S}^{(K+1)}-E\,I^{(K+1)}\right )}
 \,
 \ee
or, due to Eq.~(\ref{eurfund}),
 \be
 {\cal G}(E)=\frac{\det \left ({\cal S}^{(K)}-E\right )}
 {\det \left ( {\cal R}^{(K)}-E\,I^{(K)}\right )}
 \,
 \ee
i.e., equivalently,
 \be
 \det\,
  \left[ \begin {array}{ccccc}
  a_0 - E_{}-{\cal G}(E)&\rho_0&0&\ldots&0
    \\
     1&a_1 - E_{}&\rho_1&\ddots&\vdots
    \\
   0&1&a_2 - E_{}&\ddots&0
   \\
   \vdots&\ddots&\ddots&\ddots&\rho_{K-1}
   \\0&
   \ldots&0&1&a_K - E_{}
   \\
 \end {array} \right]
 =0\,.
 \label{Kdkit}
 \ee
On these grounds,
relation (\ref{Kdkit}) should be perceived as
a polynomial algebraic equation connecting the
$(K+1)-$plet of the unknown values of $a_j$s with the
$K-$plet of the unknown values of $\rho_k$s
at an arbitrary value of $E$.
In a way sampled, at $K=1$,
by Eq.~(\ref{ome}) above,
one is entitled to expect that in general, the choice of a $(2K+1)-$plet
of some independent energy values $E=E_\alpha$
(where, say, $\alpha=0,1,\ldots, 2K$)
would convert Eq.~(\ref{Kdkit}) into an implicit
definition of the whole
to-be-reconstructed Hamiltonian matrix $H^{(M+K)}$.

\begin{lemma}
\label{lemnis}
At any $K$
the reconstruction-providing set of equations has the following form,
 \be
 \det\,
 \left\{ \left[ \begin {array}{ccccc}
  a_0-{\cal G}(E_\alpha)&\rho_0&0
 &\ldots&0
   \\
     1&a_1&\rho_1
 &\ddots&\vdots
   \\
   0&1&a_2&\ddots
 &0
   \\
 \vdots&\ddots&\ddots&\ddots&\rho_{K-1}
    \\
  0&\ldots&0&1&a_{K}
    \\
 \end {array} \right]
 - E_\alpha\,I
 \right \}=0\,,
 \ \ \ \
 \alpha=0,1,\ldots,2K\,.%\ \ \ \ \ \ \clubsuit
 \label{dkit}
 \ee
 \end{lemma}
On these grounds,
we will now pay attention to the
entirely explicit determination of $H^{(M+K)}$
at $K=2$ and at $K=3$.

\subsection{Explicit construction at $K=2$}

At $K=2$, the five unknown quantities
$\left\{a_{{0}}, a_{{1}},a_{{2}},{\it \rho}_{{0}},
{\it \rho}_{{1}} \right\}$
forming the to-be-reconstructed matrix
 \be
 {\cal S}_{}^{(3)}=
  \left[ \begin {array}{ccc}
  a_0&\rho_0&0
   \\
     1&a_1&\rho_1
   \\
   0&1&a_2
   \\
 \end {array} \right]\,.
 \label{2spodkit}
 \ee
have to be defined
by
the quintuplet of equations
 $$
 -{E_{\alpha}}^{3}+ \left( a_{{2}}+a_{{0}}+a_{{1}}-G_{\alpha} \right)
 E_{\alpha}^{2}+ \left(  \left( a_{{1}}+a_{{2}} \right) G_{\alpha}-a_{{0}}a_{{1}}-
 a_{{1}}a_{{2}}-a_{{0}}a_{{2}}+{\it \rho}_{{0}}+{\it \rho}_{{1}} \right) E_{\alpha}+
 $$
 \be
 +\left( -a_{{1}}a_{{2}}+{\it \rho}_{{1}} \right) G_{\alpha}+a_{{0}}a_
 {{1}}a_{{2}}-{\it \rho}_{{0}}a_{{2}}-{\it \rho}_{{1}}a_{{0}}=0\,,\ \ \ \ \
 \alpha=0,1,2,3,4\,.
 \label{pata}
 \ee
The formal linearization of these equations
proceeds via the same compactification of the notation as above.
We have to consider, first, the linear system
 \be
   {E_{\alpha}}^{2}x_{{1}} +E_{\alpha}\,x_{{2}}
  +x_{{3}}
 - G_{\alpha} \, E_{\alpha}\,y_{{1}}
  - G_{\alpha}\,y_2
  ={E_{\alpha}}^{3}+G_{\alpha} \, {E_{\alpha}}
  \,,\ \ \ \ \
 \alpha=0,1,2,3,4\,
 \label{gopata}
 \ee
in which we abbreviated
 \ben
 a_{{0}}+a_{{1}}+a_{{2}}=x_1\,,\ \ \
 -a_{{0}}a_{{1}}+\rho_{{0}}-a_{{1}}a_{{2}}-a_{{0}}a_{{2}}+\rho_
{{1}}=x_2
 \,,\ \ \
 a_{{0}}a_{{1}}a_{{2}}-\rho_{{0}}a_{{2}}-a_{{0}}\rho_{{1}}=x_3\,,
 %\label{jeho}
 \een
 \be
 %\,,\ \ \
 -a_{{1}}-a_{{2}}=y_1
 \,,\ \ \
 a_{{1}}a_{{2}}-\rho_{{1}}=y_2\,.
 \label{wa}
 \ee
The solution
of Eq.~(\ref{gopata})
is obtainable via
a very routine five-by-five matrix inversion.
The input information as
initially
carried by the ten empirical parameters
$\left\{ E_0, E_1,\ldots,E_4,G_0,G_1,\ldots,G_4\right\}$
becomes now converted into a more compact input dynamical information
set
$\left\{ x_1, x_2,x_3,y_1,y_2\right\}$ with the mere five elements.

In a decisive step of the procedure we now
have to
reconstruct the quintuplet of the unknown matrix elements
$\left\{a_{{0}}, a_{{1}},a_{{2}},{\it \rho}_{{0}},
{\it \rho}_{{1}} \right\}$, i.e., we have
to solve
the coupled nonlinear algebraic
set of five equations (\ref{wa}).
For the purposes of an amended
insight in the structure of these equations
we define the following auxiliary quintuplet of polynomials
 $$
 a_{{0}}+a_{{1}}+a_{{2}}=p_1\,,\ \ \
 -a_{{0}}a_{{1}}+\rho_{{0}}-a_{{1}}a_{{2}}
 -a_{{0}}a_{{2}}+\rho_
{{1}}=p_2
 \,,\ \ \
 a_{{0}}a_{{1}}a_{{2}}-\rho_{{0}}a_{{2}}-a_{{0}}\rho_{{1}}
 =p_3\,,
 $$
 \be
 %\ \ \
 -a_{{1}}-a_{{2}}=q_1
 \,,\ \ \
 a_{{1}}a_{{2}}-\rho_{{1}}=q_2\,.
 \label{sand}
 \ee
After such an abbreviation, the solution of
our inverse $K=2$ problem becomes reduced to
the solution of the following set
 $$
  p_1(a_0,a_1,a_2)=x_1\,,\ \ \
  p_2(a_0,a_1,a_2,\rho_0,\rho_1)=x_2\,,\ \ \
  p_3(a_0,a_1,a_2,\rho_0,\rho_1)=x_3
  %\,,\ \ \
  $$
  \be
 \label{usand}
  q_1(a_1,a_2)=y_1\,,\ \ \
  q_2(a_1,a_2,\rho_1)=y_2\,.
  \ee
The solution of these
coupled polynomial equations
can already hardly be found without
computer-assisted symbolic manipulations.
Fortunately,
the standard Gr\"{o}bner elimination technique
appears applicable and yields
the closed-form result,
 \be
  a_{{0}}=y_{{1}}+x_{{1}},
   \label{ros1x} \ee
   \be
   \rho_{{0}}=-{y_{{1}}}^{2}-y_{{1}}
x_{{1}}+y_{{2}}+x_{{2}},
 \label{ros2x} \ee
 \be
 a_{{1}}=-{
\frac {-2\,y_{{2}}y_{{1}}-y_{{2}}x_{{1}}+x_{{3}}+{y_{{1}}}^{3}+{y_{{1}
}}^{2}x_{{1}}-y_{{1}}x_{{2}}}{{y_{{1}}}^{2}+y_{{1}}x_{{1}}-y_{{2}}-x_{
{2}}}},
 \label{ros3x} \ee
 \be
 \rho_{{1}}=-{\frac {C}{ \left( {y_{
{1}}}^{2}+y_{{1}}x_{{1}}-y_{{2}}-x_{{2}} \right) ^{2}}},\ \ \ \
 %\label{ros4x} \ee
% \be
 a_{{2}}=-{\frac {y_{{2}}y_{{1}}+y_{{2}}x_{{1}}
-x_{{3}}}{{y_{{1}}}^{2}+y_{{1}}x_{{1}}-y_{{2}}-x_{{2}}}}\,
 \label{ros5x} \ee
where we abbreviated
 \ben
 C={y_{{2}}}^{2}y_{{1}}x_{{1}}-3\,y_{{2}}y
_{{1}}x_{{3}}+{y_{{2}}}^{2}{x_{{1}}}^{2}-2\,y_{{2}}x_{{1}}x_{{3}}+{x_{
{3}}}^{2}+
 %\label{ros4bx}
 \een
 \ben
 +{y_{{1}}}^{3}x_{{3}}+{y_{{1}}}^{2}x_{{1}}x_{{3}}-{y_{{1}}}^{
2}x_{{2}}y_{{2}}-y_{{1}}x_{{2}}y_{{2}}x_{{1}}-y_{{1}}x_{{2}}x_{{3}}+{y
_{{2}}}^{3}+2\,{y_{{2}}}^{2}x_{{2}}+y_{{2}}{x_{{2}}}^{2}\,.
 %\label{ros4cx}
 \een
This result
encouraged us to make the following tentative prediction.

\begin{conj}
\label{conjone}
At any positive integer $K$,
all of the to-be-reconstructed matrix elements
of the full-space Hamiltonian can be expressed, in closed form, as
ratios of two polynomial functions of the
dynamical-input parameters $E_\alpha$ and  $G_\alpha$
(or, if you wish, of
$x_m$ and $y_n$).
\end{conj}

%$K\leq 3$ Motivation:

\subsection{Explicit Gr\"{o}bner-elimination solution at $K= 3$\label{excess}}

In the first nontrivial test of the acceptability of
Conjecture \ref{conjone} we decided to construct
the explicit and
exact rational-function
solution of our inverse problem at $K=3$
using a computer-supported symbolic-manipulation software.
The Conjecture passed the test. The result proved easily
obtainable via the
Gr\"{o}bner-basis-based algorithm again. The reconstruction of
 \be
 {\cal S}_{}^{(4)}=
  \left[ \begin {array}{cccc}
  a_0&\rho_0&0&0
   \\
     1&a_1&\rho_1&0
   \\
   0&1&a_2&\rho_2
   \\
   0&0&1&a_{3}
    \\
 \end {array} \right]\,
 \label{3spodkit}
 \ee
did work and proceeded again
as a more or less routine solution of the
seven coupled polynomial equations (\ref{dkit}).
% \be
% \det\,
%  \left[ \begin {array}{cccc}
%  a_0 - E_\alpha-{\cal G}(E_\alpha)&\rho_0&0&0
%    \\
%     1&a_1 - E_\alpha&\rho_1&0
%    \\
%   0&1&a_2 - E_\alpha&\rho_2
%   \\
%   0&0&1&a_3 - E_\alpha
%   \\
% \end {array} \right]
% =0\,,
% \ \ \ \
% \alpha=0,1,2,3,4,5,6\,.
% \label{3dkit}
% \ee
Nevertheless,
up to the following three exceptions
 $$
 a_{{0}}=-y_{{1}}-x_{{1}},
 $$
 $$
 \rho_{{0}}=-{y_{{1}}}^{2}-y_{{1}}x_{{1}}-y_{{2}}-x_{{2}},
 $$
 $$
 a_{{1}}={\frac {2\,y_{{1}}y_{{2}}+x_{{1}}y_{{2}}+x_{{3}}
 +{y_{{1}}}^{3}+{y_{{1}}}^{2}x_{{1}}+y_{{1}}x_{{2}}+y_{{3}}}{{y_{{1}}}^{2}
 +y_{{1}}x_{{1}}+y_{{2}}+x_{{2}}}},
 $$
the resulting formulae appeared excessively long.
Their structure was far from transparent.
For example, the full printed form of the
expression for
 $$
 \rho_{{1}}={\frac
{-{y_{{1}}}^{3}x_{{3}}+{x_{{1}}}^{2}y_{{3}}y_{{1}}+x_{{1}}y_{{2}}y_{{1
}}x_{{2}}+{y_{{2}}}^{3}+\ldots
%
%x_{{2}}x_{{4}}+x_{{4}}{y_{{1}}}^{2}+2\,x_{{2}}
%{y_{{2}}}^{2}+{x_{{2}}}^{2}y_{{2}}-x_{{1}}{y_{{2}}}^{2}y_{{1}}+{y_{{1}
%}}^{2}y_{{2}}x_{{2}}+x_{{4}}y_{{1}}x_{{1}}-y_{{1}}x_{{2}}x_{{3}}+y_{{3
%}}{y_{{1}}}^{2}x_{{1}}-x_{{1}}y_{{2}}y_{{3}}+x_{{2}}y_{{3}}x_{{1}}-x_{
%{3}}{y_{{1}}}^{2}x_{{1}}-3\,y_{{1}}y_{{2}}x_{{3}}-2\,y_{{1}}y_{{2}}y_{
%{3}}-{x_{{1}}}^{2}{y_{{2}}}^{2}-{x_{{3}}}^{2}-2\,x_{{1}}y_{{2}}x_{{3}}
%-{y_{{3}}}^{2}
%
+x_{{4}}y_{{2}}-2\,x_{{3}}y_{{3}}}{ \left( {y_{{1}}}^{2}
+y_{{1}}x_{{1}}+y_{{2}}+x_{{2}} \right) ^{2}}},
 $$
did not fit in one standard page.
Similarly, we obtained a comparatively long formula for
 \be
 \label{adve}
 a_{{2}}={\frac {x_{{1}}
{y_{{3}}}^{2}y_{{2}}+4\,y_{{1}}x_{{1}}{y_{{2}}}^{2}y_{{3}}+2\,{y_{{1}}
}^{3}x_{{1}}x_{{2}}y_{{3}}+\ldots
+{y_{{3}}}^{3}-4\,{y_{{1}}}^{2}y_{{2}}{x_{{1}}}^{2}y_{{3}}-2\,x
_{{4}}{y_{{1}}}^{2}x_{{3}}-2\,x_{{4}}y_{{1}}y_{{3}}x_{{1}}}{ \left( {y
_{{1}}}^{2}+y_{{1}}x_{{1}}+y_{{2}}+x_{{2}} \right)  \left( -{y_{{1}}}^
{3}x_{{3}}+\ldots
%
%
%{x_{{1}}}^{2}y_{{3}}y_{{1}}+x_{{1}}y_{{2}}y_{{1}}x_{{2}}+{y
%_{{2}}}^{3}+x_{{2}}x_{{4}}+x_{{4}}{y_{{1}}}^{2}+2\,x_{{2}}{y_{{2}}}^{2
%}+{x_{{2}}}^{2}y_{{2}}-x_{{1}}{y_{{2}}}^{2}y_{{1}}+{y_{{1}}}^{2}y_{{2}
%}x_{{2}}+x_{{4}}y_{{1}}x_{{1}}-y_{{1}}x_{{2}}x_{{3}}+y_{{3}}{y_{{1}}}^
%{2}x_{{1}}-x_{{1}}y_{{2}}y_{{3}}+x_{{2}}y_{{3}}x_{{1}}-x_{{3}}{y_{{1}}
%}^{2}x_{{1}}-3\,y_{{1}}y_{{2}}x_{{3}}-2\,y_{{1}}y_{{2}}y_{{3}}-{x_{{1}
%}}^{2}{y_{{2}}}^{2}-{x_{{3}}}^{2}-2\,x_{{1}}y_{{2}}x_{{3}}-{y_{{3}}}^{
%2}
%
+x_{{4}}y_{{2}}-2\,x_{{3}}y_{{3}} \right) }}\,.
 \ee
Although the full form of the latter two expressions
can only be stored in the computer, they still
represent the exact result, even though we already
did not find space
for their full display in print.

An even worse lack of space has been encountered
in the case of
our last two $K=3$ formulae for
 \be
 \label{rhodve}
 \rho_{{2}}=-{\frac {-{y_{{2
}}}^{2}{y_{{3}}}^{3}x_{{1}}+5\,{x_{{4}}}^{2}x_{{1}}y_{{2}}y_{{1}}x_{{2
}}-2\,{y_{{1}}}^{2}x_{{2}}{y_{{3}}}^{2}{x_{{1}}}^{2}y_{{2}}- \ldots
+3\,{x_{{4}}}^{2}{x_{{1}}}^{2
}y_{{3}}y_{{1}}}
{ \left( -{y_{{1}}}^{3}x_{{3}}+{x_{{1}}}^{2}y_{{3}}y_{
{1}}+x_{{1}}y_{{2}}y_{{1}}x_{{2}}+{y_{{2}}}^{3}+x_{{2}}x_{{4}}+x_{{4}}
{y_{{1}}}^{2}+2\,x_{{2}}{y_{{2}}}^{2}+\ldots
%
%{x_{{2}}}^{2}y_{{2}}-x_{{1}}{y_{
%{2}}}^{2}y_{{1}}+{y_{{1}}}^{2}y_{{2}}x_{{2}}+x_{{4}}y_{{1}}x_{{1}}-y_{
%{1}}x_{{2}}x_{{3}}+y_{{3}}{y_{{1}}}^{2}x_{{1}}-x_{{1}}y_{{2}}y_{{3}}+x
%_{{2}}y_{{3}}x_{{1}}-x_{{3}}{y_{{1}}}^{2}x_{{1}}-3\,y_{{1}}y_{{2}}x_{{
%3}}-2\,y_{{1}}y_{{2}}y_{{3}}-{x_{{1}}}^{2}{y_{{2}}}^{2}-{x_{{3}}}^{2}-
%2\,x_{{1}}y_{{2}}x_{{3}}-{y_{{3}}}^{2}
%
+x_{{4}}y_{{2}}-2\,x_{{3}}y_{{3}
} \right) ^{2}}}
 \ee
(requiring
as many as 25 lines for its full printout)
and for
 \be
 \label{atri}
 a_{{3}}=-{\frac {{y_{{2
}}}^{2}y_{{3}}+{y_{{1}}}^{2}x_{{2}}y_{{3}}-x_{{1}}y_{{3}}x_{{3}}-{x_{{
1}}}^{2}y_{{3}}y_{{2}}-x_{{1}}{y_{{3}}}^{2}-x_{{4}}{y_{{1}}}^{2}x_{{1}
}-x_{{4}}x_{{1}}y_{{2}}-2\,x_{{4}}y_{{1}}y_{{2}}- \ldots
 %x_{{4}}y_{{1}}x_{{2}}
 %-x_{{4}}x_{{3}}-x_{{4}}{y_{{1}}}^{3}+{x_{{2}}}^{2}y_{{3}}+2\,y_{{2}}x_
 %{{2}}y_{{3}}-y_{{1}}y_{{3}}x_{{3}}-y_{{1}}{y_{{3}}}^{2}+y_{{1}}x_{{1}}
 %x_{{2}}y_{{3}}-y_{{1}}x_{{1}}y_{{2}}y_{{3}}
-x_{{4}}y_{{3}}
}{-{y_{{1}}}
^{3}x_{{3}}+{x_{{1}}}^{2}y_{{3}}y_{{1}}+x_{{1}}y_{{2}}y_{{1}}x_{{2}}+{
y_{{2}}}^{3}+x_{{2}}x_{{4}}+x_{{4}}{y_{{1}}}^{2}+2\,x_{{2}}{y_{{2}}}^{
2}
%+{x_{{2}}}^{2}y_{{2}}-x_{{1}}{y_{{2}}}^{2}y_{{1}}
%
%{y_{{1}}}^{2}y_{{2
%}}x_{{2}}+x_{{4}}y_{{1}}x_{{1}}-y_{{1}}x_{{2}}x_{{3}}+y_{{3}}{y_{{1}}}
%^{2}x_{{1}}-x_{{1}}y_{{2}}y_{{3}}+x_{{2}}y_{{3}}x_{{1}}-x_{{3}}{y_{{1}
%}}^{2}x_{{1}}-3\,y_{{1}}y_{{2}}x_{{3}}-2\,y_{{1}}y_{{2}}y_{{3}}-{x_{{1
%}}}^{2}{y_{{2}}}^{2}-{x_{{3}}}^{2}-2\,x_{{1}}y_{{2}}x_{{3}}-{y_{{3}}}^
%{2}
%
+ \ldots
+x_{{4}}y_{{2}}
-2\,x_{{3}}y_{{3}}}}.
 \ee
We might only add that
the full, computer-stored form of the
latter formula is already a bit shorter again,
for the reasons which will be explained below.

\section{Compactification of formulae\label{5introduction}}

\begin{table}[h]

\caption{Reconstruction pattern.}
 \label{us1x} \vspace{.4cm}
\centering
\begin{tabular}{||c||c|c|c||}
    \hline \hline
   $K$&empirical\ data: &amended\ input& matrix elements\\
   & $E_\alpha$ and ${\cal G}(E_\alpha)$ & parameters &to be reconstructed
 \\
 \hline
 \hline
 $0$ &$\alpha=0$ & $x_1$ & $a_0$
 \\
 \hline
 $1$&$\alpha=0,1,2$ & $x_1,x_2,y_1$ & $a_0,a_1,\rho_0$
 \\
 \hline
 $2$&$\alpha=0,1,2,3,4$ & $x_1,x_2,x_3$ & $a_0,a_1,a_2$
 \\
  && $y_1,y_2$ & $\rho_0,\rho_1$
 \\
 \hline
 $3$&$\alpha=0,1,2,3,4,5,6$ & $x_1,x_2,x_3,x_4$ & $a_0,a_1,a_2,a_3$
 \\
  && $y_1,y_2,y_3$ & $\rho_0,\rho_1,\rho_2$
 \\
 \hline
 \vdots&\ldots&\ldots&\ldots\\
\hline \hline
\end{tabular}
\end{table}

 \noindent
The preceding tests made it obvious that
the
computer-generated explicit
solutions are practically useless,
mainly due to both of their complexity and excessive length.
Still, their more detailed
inspection revealed the existence of
several hidden structures which
opened the way towards the resolution of the inverse problem,
in principle at least.

The first such a hidden hint
was the possibility of the
matrix-inversion-based reduction of the
input empirical data. This feature is briefly summarized here in %the
%first two columns of
Table \ref{us1x}.
This
\textcolor{black}{change} simplified the input data
and led to
an amended formulation of the inverse problem (\ref{praka}).
%of course.

In our present paper the latter observation will be complemented
by a new, deeper insight in the algebraic
structure of the underlying nonlinear equations
mediating
an explicit constructive correspondence
between the second and the third
column in Table \ref{us1x}.
A truly essential insight will be achieved via
another amendment of the notation
in which
the emphasis will be redirected to the determinantal formulae.
Thus, first of all, we will expand our first
determinant of interest as follows,
 \be
 \det\,
  \left[ \begin {array}{ccccc}
  a_0 - \lambda&\rho_0&0&\ldots&0
    \\
     1&a_1 - \lambda&\rho_1&\ddots&\vdots
    \\
  0&1&a_2 - \lambda&\ddots&0
   \\
  \vdots&\ddots&\ddots&\ddots&\rho_{K-1}
   \\
   0&\ldots&0&1&a_K - \lambda
   \\
 \end {array} \right]
 =p_{K+1}+p_K\,\lambda+\ldots+p_1\,\lambda^K+p_0\,\lambda^{K+1}
 \,.
 \label{uKdkit}
 \ee
In parallel, we will also expand
 \be
 \det\,
  \left[ \begin {array}{ccccc}
     a_1 - \mu&\rho_1&0&\ldots&0
    \\
  1&a_2 - \mu&\rho_2&\ddots&\vdots
   \\
  0&1&a_3 - \mu&\ddots&0
   \\
  \vdots&\ddots&\ddots&\ddots&\rho_{K-1}
   \\
   0&\ldots&0&1&a_K - \mu
   \\
 \end {array} \right]
 =q_K+q_{K-1}\,\mu+\ldots+q_0\,\mu^K
 \,
 \label{vKdkit}
 \ee
and recall the direct
correspondence
of these two expansions with the second column of Table~\ref{us1x}.
Moreover, without any analogous correspondence with
the input data
we may also expand
 \be
 \det\,
  \left[ \begin {array}{ccccc}
     a_2 - \nu&\rho_2&0&\ldots&0
    \\
  1&a_3 - \nu&\rho_3&\ddots&\vdots
   \\
  0&1&a_4 - \nu&\ddots&0
   \\
  \vdots&\ddots&\ddots&\ddots&\rho_{K-1}
   \\
   0&\ldots&0&1&a_K - \nu
   \\
 \end {array} \right]
 =r_{K-1}+r_{K-2}\,\nu+\ldots+r_0\,\nu^{K-1}
 \,,
 \label{wKdkit}
 \ee
etc.

Such a notation
enables us to characterize, at any $K$,
the construction of $H^{(M+K)}$,
anew,
as based on the solution of the
determinants-related
coupled set of the
polynomial (i.e., nonlinear) algebraic equations
written in the generic form
  \be
  p_j(a_0, \ldots)=x_j\,,
  \ \ \ \ j=1,2,\ldots,K+1\,,\ \ \ \
 q_k(a_0, \ldots)=y_k\,,\ \ \ \ k=1,2,\ldots,K\,
 \label{urampa}
  \ee
and sampled, above, by Eq.~(\ref{toma}) at $K=1$,
and by Eq.~(\ref{wa}) at $K=2$.

\begin{table}[h]
\caption{A sample of consequences of our sign convention
 $p_0=(-1)^{K+1}$ and $q_0=(-1)^{K}$.}
 \label{vs1x} \vspace{.4cm}
\centering
\begin{tabular}{||c||c|c|c||c|c|c||}
    \hline \hline
    $K$&$p_1(a_0,a_1, \ldots,a_K)$ &$p_2(a_0, \ldots,a_K,\rho_0,\rho_1)$&\ldots&$q_1(a_1,a_2, \ldots,a_K)$&$q_2(a_1, \ldots,a_K,\rho_1)$&\ldots\\
 \hline
 \hline
 $0$ & $a_0$ & & &  &&
 \\
 %\hline
 $1$ & $-a_0-a_1$ & $-\rho_0+a_0a_1$ && $a_1$&&
 \\
 %\hline
 $2$ & $a_0+a_1+a_2$ & $\rho_0-a_0a_1+\ldots$&\ldots&
 $-a_1-a_2$ & $-\rho_1+a_1a_2$&
 \\
 $3$ & $-a_0-a_1-a_2-a_3$ &  $-\rho_0+a_0a_1-\ldots$ &\ldots&
 $a_1+a_2+a_3$ & $\rho_1-a_1a_2+\ldots$&\ldots
 \\
 %\hline
 \ldots&\ldots&\ldots&\ldots&\ldots&\ldots&\ldots\\
\hline \hline
\end{tabular}
\end{table}

In the new notation, it is only necessary to keep in mind
the sign conventions as
summarized in Table \ref{vs1x}
and reflecting
the natural
but important
rules $p_0=r_0=(-1)^{K+1}$ (etc)
and
$q_0=(-1)^{K}$ (etc).

%\subsection{Solutions $a_0$ and  $\rho_0$ at all $K$}

\subsection{The case of $K=2$ revisited}

%{The language of determinants}
%
%.

At $K=2$,
our coupled set (\ref{dkit})
of the five polynomial
equations
 \be
 \det\,
  \left[ \begin {array}{ccc}
  a_0 - E_\alpha-{\cal G}(E_\alpha)&\rho_0&0
    \\
     1&a_1 - E_\alpha&\rho_1
    \\
   0&1&a_2 - E_\alpha
   \\
 \end {array} \right]
 =0\,,
 \ \ \ \
 \alpha=0,1,2,3,4\,
 \label{2dkit}
 \ee
can be rewritten in its
equivalent form of Eq.~(\ref{pata})
which can be formally re-read as a linear set
which compactifies the input information
about dynamics (cf.
the general form of this compactification
as outlined in Table \ref{us1x}).
Relations (\ref{wa}) {\it alias\,} (\ref{usand})
(i.e., our fundamental polynomial $K=2$ equations rewritten
in a simplified form
using abbreviations (\ref{sand}))
can be then given an alternative
derivation via
the power-series expansions %of the following two determinants,
 \be
 \det\,
  \left[ \begin {array}{ccc}
  a_0 - \lambda&\rho_0&0
    \\
     1&a_1 - \lambda&\rho_1
    \\
   0&1&a_2 - \lambda
   \\
 \end {array} \right]
 =p_3+p_2\,\lambda+p_1\,\lambda^2+p_0\,\lambda^3
 \,,
 \label{u2dkit}
 \ee
 \be
 \det\,
  \left[ \begin {array}{cc}
       a_1 - \mu&\rho_1
    \\
   1&a_2 - \mu
   \\
 \end {array} \right]
 =q_2+q_1\,\mu+q_0\,\mu^2
 \,.
 \label{v2dkit}
 \ee
This definition yields $p_0
=-1$ and $q_0=1$ in a way compatible with our choice of signs
in
Eq.~(\ref{sand}).
\textcolor{black}{Moreover,
the addition of
$p_2(a_0, \ldots,a_2,\rho_0,\rho_1)$ and
$q_2(a_1, \ldots,a_2,\rho_1)$
eliminates $\rho_1$ and
yields, therefore, $$
 \rho_{{0}}=y_{{2}}-{y_{{1}}}^{2}-y_{{1}}x_{{1}}+x_{{2}}
 =x_2+y_2-y_1(x_1+y_1)=x_2+y_2-a_0\,y_1\,.
 $$}
An insight gained by such a change of perspective
opened in fact the path towards our constructive approach
and solutions of our inverse problem at $K>2$.
In fact, the necessity of such a change of perspective
became more than obvious as early as at $K=3$.

\subsection{The case of $K=3$ revisited}

Let us still consider just
$K=3$. For this choice
we have to determine as many as
seven unknown quantities
 $$
 \left\{a_{{0}}, a_{{1}},a_{{2}},a_{{3}},
 {\it \rho}_{{0}},{\it \rho}_{{1}},{\it \rho}_{{2}} \right\}\,.
 $$
The computer-generated answer
as outlined in preceding subsection was just the very long
closed-form solution
of
seven coupled polynomial equations (\ref{dkit}).
Now we intend to show that this result can be rearranged
and represented in compact form.

In a preparatory step
the routine
matrix-inversion
solution of the $K=3$ analogue of
the $K=2$ Eq.~(\ref{gopata})
is to be performed again.
This
enables us to replace
the initial dynamical parameters
$$\left\{ E_0, E_1,\ldots,E_6,{\cal G}(E_0),
{\cal G}(E_1),\ldots,{\cal G}(E_6)\right\}$$
by the new set
of the encoded experimental input data
$\left\{ x_1, x_2,x_3,x_4,y_1,y_2,y_3\right\}$.

In the next step we can proceed
in parallel with our
previous $K=2$ analysis
and define seven polynomial functions
$p_m$ and $q_n$
of our seven unknowns $a_j$ and $\rho_k$
via formulae
 \be
 \det\,
  \left[ \begin {array}{cccc}
  a_0 - \lambda&\rho_0&0&0
    \\
     1&a_1 - \lambda&\rho_1&0
    \\
   0&1&a_2 - \lambda&\rho_2
   \\
   0&0&1&a_3 - \lambda
   \\
 \end {array} \right]
 =p_4+p_3\,\lambda+p_2\,\lambda^2+p_1\,\lambda^3+p_0\,\lambda^4
 \,
 \label{u3dkit}
 \ee
(where $p_0=1$) and
 \be
 \det\,
  \left[ \begin {array}{ccc}
       a_1 - \mu&\rho_1&0
    \\
   1&a_2 - \mu&\rho_2
   \\
   0&1&a_3 - \mu
   \\
 \end {array} \right]
 =q_3+q_2\,\mu+q_1\,\mu^2+q_0\,\mu^3
 \,
 \label{v3dkit}
 \ee
(with $q_0=-1$).

In a re-derivation of the above-outlined $K=3$ result
in a more compact recursive form
we have to solve equations $p_m=x_m$ and $q_n=y_n$
at all $m$ and $n$.
In other words,
we have to invert the nonlinear algebraic mapping
$\left \{a_j,\rho_k
\right \} \to
\left \{p_m,q_n
\right \}
$.
The strategy
is as follows: We add
polynomials $p_1$ and $q_1$
and eliminate
 $
 a_{{0}}=-y_{{1}}-x_{{1}}\,
 $ while noticing that only
the sign has changed in comparison with the similar $K=2$ result.
In this step, as we see, the validity of the $K=3$ item in
Table~\ref{os1x} is confirmed.

Next, guided by the analogy,
the sum of the quadratic polynomials $p_2$ and $q_2$
yields the formula
 $$
 \rho_{{0}}
 =-x_2-y_2-y_1(x_1+y_1)=-x_2-y_2+a_0\,y_1
 \,
 $$
which can be found listed in Table~\ref{os2x} below.

The sum $p_3+q_3$ of the two cubic polynomials
has a less obvious structure.
A key to its analysis has been found in the discovery that it
contains five terms forming single expression $a_0q_2$.
This
led to another closed polynomial formula for the product
of the already known $\rho_0$ with the to-be-reconstructed
matrix element $a_1$,
 $$
 -\rho_0\,a_1=
 2\,y_{{1}}y_{{2}}+y_{{1}}x_{{2}}+{y_{{1}}}^{2}x_{{1}}+x_{{3}}+y_{{3}}+
 x_{{1}}y_{{2}}+{y_{{1}}}^{3}=
 $$
 $$
 =x_3+y_3+(x_2+y_2)\,y_1+(x_1+y_1)\,(y_2+y_1^2)
 =x_3+y_3-a_0\,y_2-\rho_0\,y_1
 \,.
 $$
Along the same lines
one can proceed to a completion of the task, with the
details left to the readers.

% extrapolative/ions
%$H_{M+jM+k}^{(M+K)}$ and
% at the first few $j$ and $k$ and at any~$K<\infty$

\section{Solutions valid
at all $K<\infty$\label{5Bintroduction}}

A computer-assisted implementation of
the Gr\"{o}bner elimination technique beyond $K=3$
remains routine.
We found that
the exact matrix elements of $H^{(M+K)}$
are still obtained
in a
rational-function form. Not too surprisingly, nevertheless,
such an explicit form of the result
(as sampled, above, at $K=3$) is
hardly transparent. It is increasingly
long and, in particular, it is far too long for being
presented in print.

In the light of this computation-mediated
experience
the main message delivered by our present paper
can be formulated as a conjecture that
the
above-described
recursive compactification of the $K=2$ and
$K=3$ formulae
has a systematic
extension to any finite increase of dimension $K=4,5,\ldots\,$.
In this sense, the Feshbach's inverse problem
can be expected to be
solvable at any~$K$.

The expectation of validity of such a conjecture
will be now
supported
by a few samples of its constructive verification.

\subsection{The final compact recursive form of
$a_0=H_{MM}^{(M+K)}$ (all $K \geq 0$)}

%first, trivial task: The list of
%explicit expressions for  at any $K$

%\subsection{Formulae}

%Let us remind the readers that a
%
%
%\subsubsection{The $K-$dependent form of
%$a_0=H_{MM}^{(M+K)}\,$}

During our transition to
the reconstructions of $H^{(M+K)}$ using a
variable $K$
we introduced the convention which reads
$p_0=(-1)^{K+1}$ and $q_0=(-1)^{K}$.
A sample of some immediate consequences of such a convention
is given in
Table \ref{vs1x} above.
An inspection of this Table
reveals that
the
linearity of
functions $p_1=p_1(a_0,a_1, \ldots,a_K)$
and $q_1=q_1(a_1,a_2, \ldots,a_K)$
enables us to deduce our first universal
though still truly elementary
result
 \be
 a_{{0}}=(-1)^K(x_{{1}}+y_{{1}})\,.
 \ee
This formula holds at
{\em any\,}
difference of dimensions $K=N-M>0$.
This observation is also
summarized in Table \ref{os1x}
in which we notice that
besides the entirely regular
changes in the signs,
the only true
anomaly
occurs at $K=0$.
This is
due to the fact that the input datum $y_1$
does not exist at $K=0$. We may simply put $y_1=0$ in such a case.

\begin{table}[h]

\caption{Oscillatory $K-$dependence
of the reconstructed matrix element
$a_0=H_{MM}^{(M+K)}$.}
 \label{os1x} \vspace{.4cm}
\centering
\begin{tabular}{||c||c||c|c|c|c|c||}
    \hline \hline
   $K$
& 0& 1 & 2
&3&4& \ldots
 \\
 \hline
$a_0$& $x_1$& $-x_1-y_1$ & $x_1+y_1$
&$-x_1-y_1$ & $x_1+y_1$& \ldots
 \\
\hline \hline
\end{tabular}
\end{table}

\subsection{The final compact recursive form of
$\rho_0=H_{MM+1}^{(M+K)}H_{M+1M}^{(M+K)}$ (all $K \geq 1$)}

%{Another easy task: Expression for $\rho_0$ }At $K=2$ we find

At any $K$
the sum of the two quadratic polynomials
$p_2(a_0, \ldots,a_K,\rho_0,\rho_1)$ and
$q_2(a_1, \ldots,a_K,\rho_1)$
does not contain $\rho_1$ and
yields, therefore,
a compact expression for \textcolor{black}{$\rho_{{0}}$}.
The validity of this result can be re-verified
also by the explicit computer-based
Gr\"{o}bner-elimination calculations.
At all $K$ the resulting formulae
may be found
displayed in Table \ref{os2x}.
The idea of their recursive representation makes them
remarkably
compact.

%\subsection{Compactifications of $H_{M+1,M+1}$ and of $\rho_1$ at all $K$}

\begin{table}[h]
\caption{Recursively defined off-diagonal quantities
$\rho_0=H_{MM+1}^{(M+K)}H_{M+1M}^{(M+K)}$.
}
 \label{os2x} \vspace{.4cm}
\centering
\begin{tabular}{||c||c||c|c|c|c|c||}
    \hline \hline
   $K$
& 1 & 2
&3& 4& 5&\ldots
 \\
 \hline
$\rho_0$& $-x_2+a_0\,y_1$& $x_2+y_2-a_0\,y_1$ & $-x_2-y_2+a_0\,y_1$
&$x_2+y_2-a_0\,y_1$& $-x_2-y_2+a_0\,y_1$& \ldots
 \\
\hline \hline
\end{tabular}
\end{table}

An anomaly only occurs at $K=1$ where we have to set $y_2=0$.
Via a (much easier) backward insertion
these extrapolations were verified to
hold  at a long sequence of $K$s.
The existence of these pragmatic and empirical
results also opens the way towards their
(still fairly elementary) rigorous proof
(by mathematical induction,
using the elementary properties of the determinants).

\subsection{The final compact recursive form of
$a_1=H_{M+1M+1}^{(M+K)}$ (all $K \geq 1$)}

 \noindent
From the latter two Tables
\ref{os1x} and
\ref{os2x}
one can extract an expectation that
for any matrix element of
submatrix ${\cal S}^{(K+1)}=Q^{(+)}H^{(M+K)}Q^{(+)}$ of the
to-be-reconstructed Hamiltonian $H^{(M+K)}$
there exists a critical value of $K$ beyond which
its recursive definition
remains basically (i.e., up to some signs) the same,
with the length and
complexity of the formula which
is not growing anymore.

\begin{table}[h]

\caption{Reconstructed matrix elements $a_1=H_{M+1,M+1}^{(M+K)}$.}
 \label{os3x} \vspace{.4cm}
\centering
\begin{tabular}{||c||c||}
    \hline \hline
$K$&  $a_1$ \\
    \hline \hline
 1&
 $y_1 $
 \\
 2&
 $-y_1+(x_3-a_0\,y_2)/\rho_0$
 \\
    \hline \hline
 3&
 $y_1-(x_3+y_3-a_0\,y_2)/\rho_0$
 \\
 4&
 $-y_1+(x_3+y_3-a_0\,y_2)/\rho_0$
 \\
 5&
 $y_1-(x_3+y_3-a_0\,y_2)/\rho_0$
 \\
 6&
 $-y_1+(x_3+y_3-a_0\,y_2)/\rho_0$
 \\
 %7&
% $y_1-(x_3+y_3-a_0\,y_2)/\rho_0$
% \\
% 8&
% $-y_1+(x_3+y_3-a_0\,y_2)/\rho_0$
% \\
 \vdots&\vdots
 \\
\hline \hline
\end{tabular}
\end{table}

A good test and partial verification
of such a conjecture
could be -- and has been -- provided by the
expressions for $a_1$ evaluated at
a series of $K$s.
Unfortunately,
the resulting
computer-generated sequence of the
explicit expressions
representing the next reconstructed element $a_1$
proved increasingly long.
What had to follow was a trial-and-error
rearrangement of these formulae ``by hand''.
As a result we
obtained
the
sequence of the
unexpectedly
compact formulae as displayed
in Table~\ref{os3x}.

\section{Stabilization of recurrences beyond the small $K$s\label{5Cintroduction}}

In a brief comment,
let us only return to the
case of $K=3$ and to the related
definitions~(\ref{adve}),
(\ref{rhodve}) and (\ref{atri}) of
$a_2$, $\rho_2$ and $a_3$, respectively.
Although all of these computer-generated formulae
are exact, their conventional printed form
happened to be excessively
long. As we pointed out,
we could only store them
in the
computer's memory.
This, naturally,
lowered their applicability appeal.
So we turned our attention to the
constructions of
the recursive formulae.
Our efforts were rewarded: We found that
these formulae became unexpectedly compact
and, at the larger $K$s, practically
(i.e., up to the sign changes) $K-$independent.

\subsection{Final recursive form of
$\rho_1=H_{M+1M+2}^{(M+K)}H_{M+2M+1}^{(M+K)}$ ($K \geq 2$)}
%and of $a_2=H_{M+2M+2}^{(M+K)}$}

%{Expression for $\rho_1$ at any $K$}

As long as the
highest, quartic polynomial $p_4$
has no partner $q_4$
at $K=3$,
our above-explained knowledge of the
explicit rational-function $K=3$ result
does not offer any obvious way of
extracting a formula for $\rho_1$.
Fortunately, the way towards such an extraction
has been found in the
subtraction of products $a_0\,q_3$ and $\rho_0\,q_2$.
This
converted the underlying inverse-problem equation $p_4=x_4$
(emerging as an $K=3$ analogue of
its $K=2$ predecessor~(\ref{usand}))
into our ultimate rational-function formula
 $$
 \rho_1=
 y_2+a_1\,y_1-a_1^2
 +\frac{a_0\,y_3-x_4}{\rho_0}\,.
 $$
The knowledge of this special $K=3$ formula opened the way
towards its extension
to the larger $K$s, with the calculated
(and, by the backwards insertion, reconfirmed)
results listed in Table~\ref{os4x}.

\begin{table}[h]

\caption{Reconstructed products
$\rho_1=H_{M+1,M+2}^{(M+K)}H_{M+2,M+1}^{(M+K)}$.}
 \label{os4x} \vspace{.4cm}
\centering
\begin{tabular}{||c||c||}
    \hline \hline
$K$&  $\rho_1$ \\
    \hline \hline
 2&
 $-a_1^2-y_2-a_1\,y_1 $
 \\
 3&
 $-a_1^2+y_2+a_1\,y_1-(x_4-a_0\,y_3)/\rho_0$\\
 \hline
 \hline
 4&
 $-a_1^2-y_2-a_1\,y_1+(x_4+y_4-a_0\,y_3)/\rho_0$\\
 5&
 $-a_1^2+y_2+a_1\,y_1-(x_4+y_4-a_0\,y_3)/\rho_0$\\
 6&
 $-a_1^2-y_2-a_1\,y_1+(x_4+y_4-a_0\,y_3)/\rho_0$\\
 7&
 $-a_1^2+y_2+a_1\,y_1-(x_4+y_4-a_0\,y_3)/\rho_0$\\
 8&
 $-a_1^2-y_2-a_1\,y_1+(x_4+y_4-a_0\,y_3)/\rho_0$\\
 \vdots&\vdots
 \\
\hline \hline
\end{tabular}
\end{table}

All of the constructive results as displayed in
Tables \ref{os1x} -- \ref{os4x}
may be interpreted as a
support of their following extrapolation.

\begin{conj}
\label{fohy}
In recurrent manner,
every
to-be-reconstructed matrix element $H_{M+i,M+j}^{(M+K)}$ of
Hamiltonian (\ref{hejkitie}) (or, more precisely, (\ref{fkitie}))
with subscript $(i,j) = (0,0), (0,1), (1, 1),
(1,2), \ldots ,(K,K)$
can be defined
as a comparatively compact $i-$ and $j-$dependent
rational function
of all of its predecessors $H_{M+i',M+j'}^{(M+K)}$
with $i'<i$ or $j'<j$.
\end{conj}

\subsection{Solutions at $K\geq K_{stab}$.}

In connection with Conjecture \ref{fohy}
we decided to test the hypothesis
by the next-step reconstruction
of the matrix element $a_2$ at all $K$.
The result is displayed in
our last Table~\ref{os5x}.
An independent, additional
reason for the presentation of the latter Table
can be also seen
in
the emergence of a partial-fraction structure
of the formulae.
All of the technical details
of the derivation of this result
are omitted here because their
description would require
a disproportionate prolongation of the text.

\begin{table}[h]
\caption{Reconstructed matrix elements $a_2=H_{M+2M+2}^{(M+K)}$.}
%=H_{M+2,M+2}
 \label{os5x} \vspace{.4cm}
\centering
\begin{tabular}{||c||c||}
    \hline \hline
$K$&  $a_2$ \\
    \hline \hline
 2&
 $-a_1-y_1 $
 \\
 3&
 $-2\,a_1+y_1+(y_3+a_1\,y_2+a_1^2\,y_1-a_1^3)/\rho_1$\\
 4&
 $-2\,a_1-y_1-(y_3+a_1\,y_2+a_1^2\,y_1+a_1^3)/\rho_1
 +(x_5-a_0\,y_4)/(\rho_0\,\rho_1)
 $\\
 \hline \hline
 5&
 $-2\,a_1+y_1+(y_3+a_1\,y_2+a_1^2\,y_1-a_1^3)/\rho_1
 -(x_5+y_5-a_0\,y_4)/(\rho_0\,\rho_1)
 $\\
 6&
 $-2\,a_1-y_1-(y_3+a_1\,y_2+a_1^2\,y_1+a_1^3)/\rho_1
 +(x_5+y_5-a_0\,y_4)/(\rho_0\,\rho_1)
 $\\
 7&
 $-2\,a_1+y_1+(y_3+a_1\,y_2+a_1^2\,y_1-a_1^3)/\rho_1
 -(x_5+y_5-a_0\,y_4)/(\rho_0\,\rho_1)
 $\\
 8&
 $-2\,a_1-y_1-(y_3+a_1\,y_2+a_1^2\,y_1+a_1^3)/\rho_1
 +(x_5+y_5-a_0\,y_4)/(\rho_0\,\rho_1)
 $\\
 \vdots&\vdots
 \\
\hline \hline
\end{tabular}
\end{table}

An inspection of the recursive reconstruction formulae
as sampled and displayed in the quintuplet of
Tables \ref{os1x} -- \ref{os5x}
reveals the emergence of a rather unexpected
stabilization which
can tentatively be
generalized and yield the following hypothesis.

\begin{conj}
\label{conjem}
The ``length''
(i.e., the number of the components) of the forward-running recurrent
definitions of the reconstructed Hamiltonian matrix $H^{(M+K)}$ initially
grows with $K$ but it
ceases to grow
at $K= K_{stab}$,
with $K_{stab}=2m+1$
for $a_m=a_m(\rho_{m-1},a_{m-1},\rho_{m-2}, \ldots)$,
and
with $K_{stab}=2m+2$
for $\rho_m=\rho_m(a_m,\rho_{m-1},a_{m-1},\rho_{m-2}, \ldots)$.
\end{conj}

 \noindent
Such a conjecture is,
together with its above-presented constructive support,
one of our most amazing results.
The stabilization
(i.e., the existence of an upper bound  of the complexity of the recursive
formulae)
is
(or, more precisely, would be) a truly welcome guarantee that
in the realistic
applications using larger $K$s
the
whole reconstruction process can still be kept
reasonably user-friendly.

\section{Optimal construction strategy
at a fixed $K=K_0$\label{6introduction}}

%\section{Opposite-direction recursions}

%=====

In our last three or four Tables
we may notice that
at a larger and fixed,
preselected value of $K=K_0$,
the use of the forward-running recurrences
for $a_m$ and $\rho_m$
ceases to be optimal near $m \approx K_0$.
For a complete reconstruction of $H^{(M+K_{0})}$, indeed,
we need the set of the formulae up to
$m=K_{0} $.
Especially at the larger $K_0$s,
this is the
value which is perceivably
smaller than the corresponding length-stabilization
boundary
$K_{stab}=2m+1$.

In what follows we are going to outline a remedy.
At an expense of a more complicated
strategy of the construction of the solutions,
the idea of
such a remedy will be based, in essence, on a matching of the
forward-running
and backward-running recurrences
somewhere in the middle of the interval
of the subscripts, i.e., roughly,
somewhere near $m \approx K_{0}/2$.

\subsection{Revisiting the formulae at $K_0=2$}

At $K=2$ we have to solve the quintuplet
(i.e., three plus two) polynomial algebraic equations
 \ben
 a_{{0}}+a_{{1}}+a_{{2}}=x_1\,,\ \ \
 -a_{{0}}a_{{1}}+\rho_{{0}}-a_{{1}}a_{{2}}-a_{{0}}a_{{2}}+\rho_
{{1}}=x_2
 \,,\ \ \
 a_{{0}}a_{{1}}a_{{2}}-\rho_{{0}}a_{{2}}-a_{{0}}\rho_{{1}}=x_3\,,
 %\label{jeho}
 \een
 \be
 %\,,\ \ \
 -a_{{1}}-a_{{2}}=y_1
 \,,\ \ \
 a_{{1}}a_{{2}}-\rho_{{1}}=y_2\,.
 \label{ewa}
 \ee
In the forward-running direction of recurrences
we may proceed as above,
extracting $a_0$ and
$\rho_0=\rho_0(a_0)$
from the first two items of the upper triplet
of equations.

The change of the strategy
will only concern
the last two (lower-line) relations.
The first one yields $a_2=a_2(a_1)$ and,
subsequently, we get $\rho_1=\rho_1(a_1)$
from the second one.
This allows us to
treat now the value of
$a_1$ as the only unknown quantity,
determined, after all insertions, by the
third item of the upper triplet
of equations.

One can conclude that
such a form of result is really achieved via a matching of the
direct and inverted recurrence.
Still, the matching can be simplified since
the ``not yet used'' third relation on the first line of Eq.~(\ref{ewa})
can be also reread
as a direct alternative
definition of $a_2=a_2(a_1)$.
This would
yield another, simpler self-consistency constraint
imposed upon $a_1$.

Incidentally, one reveals that
the latter version of constraint acquires the form of a
linear algebraic equation which makes the matching elementary.
Leading, naturally, to the final result which is the same as above,
i.e., to the
value of $a_1$ as given in Table \ref{os3x}.

%\newpage

\subsection{Revisiting the formulae at $K_0=3$}

Starting from the reconstructions of $H^{(M+K_0)}$
with
$K_0 \geq 3$
it makes sense to return to
the polynomial representations (\ref{uKdkit}), (\ref{vKdkit})
and (\ref{wKdkit}) of the determinants.

The merits of such a change of paradigm become obvious even at
$K_0=3$. In this case we can still recall the forward-running
recurrences and treat the elements $a_0$, $\rho_0$, $a_1$ and
$\rho_1$ as known (see the respective Tables above). This enables us
to re-interpret the first four items of the $K=3$ version of
Eq.~(\ref{urampa}) as identities and to omit them as redundant. In
this way we are left with the triplet of equations
 $$
 y_1= a_{{1}}+a_{{2}}+a_{{3}}\,,
 \ \ \ \
 y_2=
  -a_{{1}}a_{{2}}+{\it \rho}_{{1}}
  -a_{{2}}a_{{3}}-a_{{1}}a_{{3}}+{\it \rho}_ {{2}}
 \,,\ \ \
 y_3= a_{{1}}a_{{2}}a_{{3}}-{\it \rho}_{{1}}a_{{3}}-a_{{1}}{\it
 \rho}_{{2}}\,.
 $$
These relations may be now treated as a source of
recurrences running in the opposite direction.
Thus, the first relation enables us to eliminate
$a_3=y_1-a_1-a_2$,
with $a_2$ re-acquiring the status of a
temporarily unknown variable in $a_3=a_3(a_2)$.
Similarly, the second relation enables us to define
$\rho_2=\rho_2(a_2)$ so that we are left with
the third relation
which finally defines the unknown $a_2$
(again, in full agreement with its $K=3$ value
as listed in Table \ref{os5x} above).

In a slightly modified arrangement of the matching
it makes sense to recall the determinantal formulae
and to complement the specific $K=3$ expansions
(\ref{u3dkit}) and (\ref{v3dkit})
by their third determinantal partner
 \be
 \det\,
  \left[ \begin {array}{cc}
     %  a_1 - \nu&\rho_1&0
%    \\   1&
   a_2 - \nu&\rho_2
   \\
   1&a_3 - \nu
   \\
 \end {array} \right]
 =r_2+r_1\,\nu+r_0\,\nu^2
 \,.
 \label{w3dkit}
 \ee
Here we have $r_0=1$ and
two abbreviations $r_1=r_1(a_2,a_3)=-a_2-a_3=a_1-y_1=r_1(a_1)$
and $r_2=a_2a_3-\rho_2$.
In such a rearrangement of the construction we
may work with
another pair of the new auxiliary variables $r_1$ and $r_2$
which are comparatively easily defined
via the determinant-related identities $x_3+y_3-a_0y_2+\rho_0r_1=0$
and $x_4-a_0y_3+\rho_0r_2=0$.

The details of
the check of such an alternative approach
to the problem of the matching
are left to the readers:
Its more systematic background
would become more important
for the
(here, skipped)
study of the more advanced
matchings and reconstructions using $K_0\geq 4$.

\section{Discussion\label{discussion}}

The description of dynamics of quantum systems is often based on
the Feshbach's and L\"{o}wdin's concept of model space.
Mathematically this means that the full-space Hamiltonian $H$
is replaced by its
model-space avatar $H_{e\!f\!f}$
which can be, in general, fairly strongly
bound-state-energy-dependent,
$H_{e\!f\!f}=H_{e\!f\!f}(E)$.
In the present paper the inverse problem has been considered.
The effective Hamiltonian was assumed known
(as, typically, a tentative
trial-and-error model which fits experimental data)
while the full-space Hamiltonian $H$
had to be reconstructed.

At the first sight, the task of reconstruction
$H_{\!\it eff}(E)\ \to\ H$
seems next-to-impossible.
Even if one only considers the
unitary quantum systems of
bound states,
the information about dynamics
as carried by the full-fledged quantum Hamiltonian $H$
seems too extensive to be reconstructed from any input
at a reasonable cost.
At the same time,
a part of the information about $H$
is only mathematical, related to the
choice of basis in Hilbert space.
One has to distinguish between
the bases given in advance (in which
$H$ is a general matrix)
and the bases which are
at least partially adapted to the Hamiltonian
(in which, typically, its representation
is a user-friendlier
sparse matrix with many elements equal to zero).
Of course, a maximum of the user-friendliness
is the diagonality
but the diagonalization is too costly
a process. For us, this led to the decision
of considering the picture of bound states in which the
Hamiltonian is merely an $N$ by $N$ matrix
with finite $N$
which is
asymptotically tridiagonal.

In such an overall setting
we assumed that the
information
about the spectrum has the form of an
energy-dependent $M$ by $M$ matrix
$H_{\!\it eff}(E)$
with $M \ll N=M+K$.
This enabled us
to reduce the general
inverse problem of reconstruction
$H_{\!\it eff}(E)\ \to\ H$
to the problem of solution
of a coupled set of $2K+1$
polynomial algebraic equations.

The solution was split in
two steps.
Firstly, we made use of the
knowledge of
the dynamics (as carried by
$H_{\!\it eff}(E)$)
and we converted this information
into a $(2K+1)-$plet of certain amended
input parameters
$x_j$ and $y_k$
(cf. Table \ref{us1x}).
This reduced the inverse problem
to a
set of nonlinear algebraic equations (cf. Eq.~(\ref{urampa}))
which had to be solved in the second step.

Initially we felt strongly discouraged by the nonlinearity
of the latter set of equations.
We even
arrived at a highly skeptical conclusion
that even the maximally simplified version of
the inverse Feshbach's problem
can hardly be resolved in closed form,
with the only exception surviving in the simplest possible scenario
in which
$K=1$ (cf. \cite{PLAI} and Lemma \ref{lemmakje1} above).
In this light, the main result of our present paper
is
a manifest constructive disproof of the latter
erroneous -- albeit not too surprising -- skepticism.

A return to optimism has been initiated
by our
computer-assisted study of the
underlying set of the coupled polynomial equations.
The first encouragement
emerged when we found
that the more or less routine application of the
Gr\"{o}bner-basis
elimination technique yields the
solutions, albeit excessively complicated, in closed form.
Initially,
the practical appeal of such a
form of solvability still appeared rather academic because
the resulting formulae
(in fact, the ratios of polynomials)
appeared prohibitively long
even at $K$ as small as three (cf. paragraph \ref{excess} above).
The reconstructed Hamiltonian
seemed to be only suitable for being
deposited in a computer's memory.
Fortunately, its subsequent scrutiny revealed that
the results
can very efficiently be made
compact
when re-written in recursive manner.

This rendered possible
the printed display of our
amended, ultimate formulae
(cf. their multiple samples as displayed
in several
illustrative Tables above).
This also formed a part of a
decisive
motivation of our return to the problem.
Another, phenomenologically
oriented part of the motivation
was provided by
a broader physical context.
During many pragmatically oriented applications
of quantum mechanics, indeed, one really has to find an
optimal balance between
a well founded theoretical model
and an agreement
of the related predictions
with some relevant, experimentally available data.

Typically, one
decides to start from a tentative choice of a
Hamiltonian and
solves the related Schr\"{o}dinger equation.
Subsequently, one has to confirm
the
phenomenological qualities
of the choice of $H$, say, via a comparison of the
calculated
low-lying bound-state energies $E^{(n)}$
with their measured values.
Naturally, beyond a few truly elementary quantum systems,
our satisfaction provided by the comparison is limited
not only by the unavoidable uncertainties in the
experimental data but also by the equally unavoidable ambiguity
of our choice of the theoretical model.

In practice, therefore,
we are forced to combine a simplification of the theory
with a certain {\it ad hoc\,} weakening of
the criteria of its
compatibility with the data.
Then, the most realistic model-building strategies
share several characteristic features.
Using the terminology of mathematics
we restrict attention, typically, to the models living
in a finite-dimensional Hilbert space.
This allows us to replace the
sophisticated formalism of functional analysis
by the language of linear algebra.
We become permitted to
replace
the original choice of the
realistic (though, presumably, also next to
prohibitively complicated) operator $H$ entering
the generic Schr\"{o}dinger bound-state problem
by
a user-friendlier truncated matrix $H^{(M)}$
and/or by
its amended effective-Hamiltonian
alternative  $H_{e\!f\!f}^{(M)}(E)$.

\section{Summary\label{susu}}

 %\noindent
In
a fundamental form of quantum
theory the full-space Hamiltonian  $H^{(N)}$ with $N \leq \infty$
is usually required to be energy-independent.
In contrast, any pragmatically motivated
initial choice of
the smaller, $M$ by $M$ matrix
$H_{e\!f\!f}^{(M)}(E)$ with $M=N-K$
(which fits the measured data)
may be expected to vary with $E$.
Such a generic form of variability
was in fact the basic idea of
our proposal of the
method of the reconstruction of $H^{(N)}$ in \cite{PLAI}.

Not too surprisingly, such a version of our proposal
was not sufficiently persuasive because
we were only able to implement the method at
$K=1$.
In this sense,
a decisive progress has only been achieved in our present
completion of the study.
First of all
we discovered
that
the underlying coupled set of
polynomial equations
seems to
lead
to the closed algebraic formulae
which define the solutions at any finite $K$.
This was formulated as a conjecture,
the plausibility of which
was supported by multiple explicit formulae.

From a purely pragmatic point of view the latter formulae
defined the full-space Hamiltonian
$H^{(M+K)}$, but this definition
appeared to be next-to-prohibitively complicated
even at $K=3$.
Fortunately, we were able to
convert this definition
into its other -- and compact -- equivalent recursive
alternative.
We revealed that
the number of terms in
such a recursive
(i.e., forward-running recurrent)
representation
of every out-of-the-model-space matrix
element $H^{(M+K)}_{M+i,M+j}$
only grows at the first few smallest indices $i$ and $j$.
Subsequently,
such a number becomes
a constant.
Even the form
of the recurrence
stabilizes and,
practically (i.e., up to some oscillating signs), it
does not change with the
further growth of the dimension $N=M+K$
of the full Hilbert space anymore.


\newpage


\begin{thebibliography}{00}


 \bibitem{Feshbach}
Feshbach H 1958 Unified theory of nuclear reactions
 {\em Ann. Phys. (NY)} {\bf 5} 357 -- 390

\bibitem{Lomb}
Hocine E, Yekken R and Lombard R J 2019
Applying quantum supersymmetry and perturbation
theory to the energy-dependent Hulthen potential
%Hocine, E; Yekken, R and Lombard, R
%Nov 2019
{\em Eur. Phys. J. PLUS} {\bf 134} 561
%(11)
%We deal with the energy-dependent Hulthen potential,
%by using the supersymmetric quantum mechanics and
%the first-order perturbation theory. We consider
%the Hulthen potential linearly dependent on the energy
%which is introduced in the coupling constant.
%We evaluate the energy eigenvalues and the corresponding reduced rad
%Volume134Issue11
%DOI10.1140/epjp/i2019-12921-6
%Article Number
%561
%Published
%NOV 2019

 \bibitem{Feshbachb}
Lombard R J and Mare\v{s} J  2009
The many-body problem with an energy-dependent confining potential
{\em Phys. Lett A} {\bf 373} 426 -- 429


 \bibitem{Feshbachc}
Alimohammadi  M and Hassanabadi H 2016
 Gamma-rigid regime of the Bohr-Mottelson
 Hamiltonian in energy-dependent approach
%Oct 2016
{\em Int. J. Modern Phys. E} {\bf 25} 1650087
%(10)
%We determine the energy spectrum and wave function for the
%Bohr-Mottelson Hamiltonian on gamma-rigid regime separately
%with the harmonic and Coulomb energy-dependent potentials.
%We study the effect of potential parameters on the energy levels
%and probability density distribution. The transition rates are determined
%in each case
%Volume25
%Issue10
%DOI10.1142/S0218301316500877

\bibitem{Loewdin}
%H. Feshbach,  {\em A unified theory of nuclear reactions II}, Ann.
%Phys. (NY) {\bf  19},  287-313, 1962;
%
L\"{o}wdin P-O 1962 Studies in Perturbation Theory IV Solution
of Eigenvalue Problem by Projection Operator Formalism
{\em J. Math. Phys.} {\bf 3} 969 -- 982
%doi:10.1063/1.1724312


\bibitem{Baha}
Bahar M K 2022
Charge-current generations and
optical specifications of Gaussian quantum dot
with energy-dependent potential
%Sep 2022
{\em Chem. Phys. Lett.} {\bf 802} 139761
%We consider the energy-dependent quantum dot with GaAs/GaAlAs
%Gaussian potential encompassment. Depending on the experimental
%or theoretical results for different types of interactions,
%the energy dependency can also be chosen as quadratic,
%fractional or any other type. The solutions of the wave
%equation for the energy dependent Gaussian quantum dot
%are performed numerically by employing the Ru
%Volume802
%DOI10.1016/j.cplett.2022.139761
%Article Number
%Published
%SEP 2022
%Early Access
%JUN 2022

\bibitem{PT}
Schulze-Halberg  A and Roy  P
2017
Pseudo-hermitian and PT-symmetric quantum systems with
energy-dependent potentials: Bound-state solutions and
energy spectra {\em Ann. Phys. (NY)} {\bf 380} 78 -- 92
%Schulze-Halberg, A and Roy, P
%May 2017
%ANNALS OF PHYSICS 380, pp.78-92
%We introduce generalized versions of complex Scarf
%and Morse-type potentials that contain energy-dependent
%parameters. PT-symmetry and pseudo-hermiticity of the
%associated quantum systems are discussed, and a modified
%orthogonality relation and pseudo-norm are constructed.
%We show that despite energy
%Volume380Page78-92
%DOI10.1016/j.aop.2017.02.014
%Published
%MAY 2017
%Indexed
%2017-05-31




\bibitem{neco}
Harko  T and Liang  S D
2019
Energy-dependent noncommutative quantum mechanics
{\em Eur. Phys. J. C} {\bf 79} 300
%
%Harko, T and Liang, SD
%Apr 3 2019
%EUROPEAN PHYSICAL JOURNAL C 79(4)
%We propose a model of dynamical noncommutative quantum mechanics in which the noncommutative strengths, describing the properties of the commutation relations of the coordinate and momenta, respectively, are arbitrary energy-dependent functions. The Schrodinger equation in the energy-dependent noncommutative algebra is derived for a two-dimensional system for an arbitrary potential. The resulti
%Volume79
%Issue4
%DOI10.1140/epjc/s10052-019-6794-4
%Article Number
%300
%Published
%APR 3 2019
%42 cits


\bibitem{DKP}
Langueur  O, Merad  M and Hamil  B
2019
DKP Equation with Energy Dependent Potentials
{\em Commun. Theor. Phys.} {\bf 71} 1069 -- 1074
%Langueur, O; Merad, M and Hamil, B
%Sep 2019
%COMMUNICATIONS IN THEORETICAL PHYSICS 71(9), pp.1069-1074
%In this work, we study the DKP equation subjected to the action
%of combined vector plus scalar energy depend on potentials
%in (1+1) dimensions space-time. The conditions of normalisation
%and continuity equation are calculated. The eigenfunctions
%and the corresponding eigenvalues are then determined.
%A numerical study is presented and the energy graphs
%for some values of the energy parameter are plotted.
%Volume71Issue9Page1069-1074
%DOI10.1088/0253-6102/71/9/1069
%Published
%SEP 2019
%relativistic equation namely Duffin-Kemmer-Petiau
%(DKP) equation, other than that of Dirac and KleinGordon, describing the dynamics of the scalar and vectorial Bosons


\bibitem{info}
Boumali A and Labidi M
2021
The Solutions on One-Dimensional Dirac Oscillator
with Energy-Dependent Potentials and
Their Effects on the Shannon and
Fisher Quantities of Quantum Information Theory
{\em J. Low Temp. Phys.} {\bf 204} 24 -- 47
%Boumali, A and Labidi, M
%Jul 2021
%JOURNAL OF LOW TEMPERATURE PHYSICS 204(1-2), pp.24-47
%Enriched Cited References
%In this paper, we focus, at first, on the exact solutions
%on the one-dimensional Dirac oscillator with
%the energy-dependent potentials. Then, the influence
%of these solutions on the Shannon entropy and Fisher
%information, well-known in quantum information,
%has been studied. In this direction, we concentrated
%on the determinati
%
%Volume204Issue1-2Page24-47
%DOI10.1007/s10909-021-02596-6
%Published
%JUL 2021
%Early Access
%MAY 2021
%Indexed
%2021-05-23


\bibitem{Hill}
Znojil M
%May 31
2004
Linear representation of energy-dependent Hamiltonians
{\em Phys. Lett. A}
{\bf 326}
%(1-2), pp.
70--76

\bibitem{Hillb}
Badanin A and Korotyaev E L 2021
Hill's operators with the potentials analytically dependent on energy
{\em J. Differential Equations} {\bf 271} 638 -- 664
% 15 2021
%JOURNAL OF DIFFERENTIAL EQUATIONS 271, pp.638-664
%We consider Schrodinger operators on the line with potentials
%that are periodic with respect to the coordinate variable
%and real analytic with respect to the energy variable.
%We prove that if the imaginary part of the potential is
%bounded in the right half-plane, then the high energy
%spectrum is real, and the corresponding asymptotics
%Volume271Page638-664
%DOI10.1016/j.jde.2020.09.016
%Published
%JAN 15 2021
%Indexed
%2021-01-05


\bibitem{Mares}
Form\'anek J, Mare\v{s} J and Lombard R J 2004
Wave equations with energy-dependent potentials
{\em  Czech J. Phys.}
{\bf 54}  289

\bibitem{Loewdinb}
Albuquerque S, V\"{o}lkel  S H and Kokkotas  K D
2024
Inverse problem in energy-dependent potentials using semiclassical methods
{\em Phys. Rev. D} {\bf 109} 096014
%
%%Albuquerque, S; Völkel, SH and Kokkotas, KD
%%May 13 2024
%PHYSICAL REVIEW D 109(9)
%
%PHYSICAL REVIEW D
%Volume109
%Issue9
%DOI10.1103/PhysRevD.109.096014
%Article Number
%096014
%Published
%MAY 13 2024
%Indexed
%2024-06-23



\bibitem{Maresb}
Lombard R J, Mare\v{s} J and Volpe C 2007
Wave equation with energy-dependent
potentials for confined systems
{\em J. Phys. G:  Nucl. Part.
Phys. }
{\bf 34} 1879 -- 1889

\bibitem{Maresc}
L\"{u}tf\"{u}oglu  B C, Ikot  A N, Karakoc  M,
Osobonye  G T, Ngiangia  A T and Bayrak  O
2021
Bound state solutions of the Klein-Gordon equation with energy-dependent potentials
{\em Mod. Phys. Lett. A} {\bf 36} 2150016
%Lütfüoglu, BC; Ikot, AN; (...); Bayrak, O
%Lütfüoglu, BC (Lutfuoglu, B. C.) [1] , [2] ; Ikot, AN (Ikot, A. N.) [3] ;
%Karakoc, M (Karakoc, M.) [1] ; Osobonye, GT (Osobonye, G. T.) [4] ;
%Ngiangia, AT (Ngiangia, A. T.) [3] ; Bayrak, O (Bayrak, O.) [1]
%Feb 10 2021
%MODERN PHYSICS LETTERS A 36(4)
%In this paper, we investigate the exact bound state solution of the Klein-Gordon
%equation for an energy-dependent Coulomb-like vector plus scalar potential energies.
%To the best of our knowledge, this problem is examined in literature with a constant
%and position dependent mass functions. As a novelty, we assume a mas
%Volume36Issue4
%DOI10.1142/S0217732321500164
%Article Number
%2150016
%Published
%FEB 10 2021

\bibitem{Marescc}
Schulze-Halberg  A
2021
Characterization of Darboux transformations for quantum
systems with quadratically energy-dependent potentials
{\em J. Math. Phys.} {\bf 62} 083504
%
%Schulze-Halberg, A
%Aug 1 2021
%JOURNAL OF MATHEMATICAL PHYSICS 62(8)
%Enriched Cited References
%We construct three classes of higher-order Darboux
%transformations for Schrodinger equations with
%quadratically energy-dependent potentials by means of
%generalized Wronskian determinants. Particular even-order
%cases reduce to the Darboux transformation for conventional
%(energy-independent) potentials. Our constructi
%Volume62Issue8
%DOI10.1063/5.0051739
%Article Number
%083504
%Published
%AUG 1 2021
%Indexed
%2021-09-19



\bibitem{Maresd}
Budaca R
2025
Particle in a spherical potential well of energy dependent depth
{\em Eur. Phys. J. PLUS}
{\bf 140} 344
%Particle in a spherical potential well of energy dependent depth
%Budaca, R
%Apr 28 2025
%EUROPEAN PHYSICAL JOURNAL PLUS 140(4)
%Volume140
%Issue4
%DOI10.1140/epjp/s13360-025-06285-1
%Article Number
%344
%Published
%APR 28 2025
%
%
%


\bibitem{Marese}
Ertugay  C, Edet C O, Ikot  A N and L\"{u}tf\"{u}oglu  B C 2025
Thermo-magnetic properties of non-relativistic particles under
the effect of energy-dependent Hellmann potential
{\em Molecular Phys.} {\bf 123} e2411327
%Ertugay, C; Edet, CO; (...); Lütfüoglu, BC
%Ertugay, C (Ertugay, C.) [1] ; Edet, CO (Edet, C. O.) [2] , [3] ;
%Ikot, AN (Ikot, A. N.) [4] , [5] ; Lütfüoglu, BC (Lutfuoglu, B. C.) [6]
%Jun 3 2025
%MOLECULAR PHYSICS 123(11)
%Volume123
%Issue11
%DOI10.1080/00268976.2024.2411327
%Article Number
%e2411327
%Published
%JUN 3 2025
%Early Access
%OCT 2024
%



\bibitem{PLAI}
Znojil M 2025
Reconstruction of full-space quantum Hamiltonian from its effective,
energy-dependent model-space projection
 {\em Phys. Lett. A} {\bf 556}  130816


{\bibitem{CF}
Znojil M 1976
The recursion method of a linear operator inversion
%J. Phys. A: Math. Gen. 9 (1976) 1-10.
{\em J. Phys. A: Math. Gen.} {\bf 9}  1 -- 10}

{\bibitem{Has}
Gesztesy F and Simon B 1997 m-Functions and inverse spectral analysis for
finite and semi-infinite Jacobi matrices.
{\em J.  Anal. Math.} {bf 73} 267 -- 297}
%(1997). https://doi.org/10.1007/BF02788147

{\bibitem{Ha}Holtz O and Tyaglov M 2012
Structured Matrices, Continued Fractions, and Root Localization of Polynomials
%Olga Holtz and Mikhail Tyaglov
{\em SIAM Review} {bf 54} 421 -- 509}
%10.1137/090781127



{\bibitem{Hashim} Yamani H A and Mouayn Z 2014
Supersymmetry of tridiagonal Hamiltonians
%Hashim A Yamani and Zouhair Mouayn
%
%Published 17 June 2014 • © 2014 IOP Publishing Ltd
{\em J. Phys. A: Math. Theor.} {\bf 47}  265203}


{\bibitem{Hash} Bagarello F, Gargano F and Roccati F
2019 Tridiagonality, supersymmetry and non self-adjoint Hamiltonians
{\em J. Phys. A: Math. Theor.} {\bf 52}   355203}

{\bibitem{PS1}
Pushnitski A and \v{S}tampach F 2024
An inverse spectral problem for non-self-adjoint Jacobi matrices
{\em Int. Math. Res. Notices} {bf 2024}  6106 -- 6139}

\end{thebibliography}
\end{document}